\documentclass[12pt]{article}

\ifx\pdfoutput\undefined
\usepackage[dvips,bookmarks]{hyperref}
\else
\usepackage{hyperref}
\fi
\hypersetup{colorlinks=false,bookmarksopen,bookmarksnumbered,citecolor=blue,
   pdfstartview=FitH}

\usepackage{latexsym}
\usepackage{amssymb,amsfonts,amsmath}
\usepackage{graphicx} 
\usepackage{indentfirst}
\usepackage{bbm}
\usepackage{amssymb}
\usepackage{verbatim}
\usepackage{amsmath, amsthm,amssymb}
\usepackage{mathrsfs}
\usepackage{hyperref}
\usepackage{amsfonts}
\usepackage{dsfont}

\parskip=\medskipamount

\newcommand {\cC}{{\cal C}}
\newcommand {\cD}{{\cal D}}

\newcommand {\cG}{{\cal G}}
\newcommand {\cH}{{\cal H}}

\newcommand {\cJ}{{\cal J}}
\newcommand {\cK}{{\cal K}}

\newcommand {\cM}{{\cal M}}
\newcommand {\cN}{{\cal N}}

\newcommand {\cR}{{\cal R}}
\newcommand {\cS}{{\cal S}}
\newcommand {\cT}{{\cal T}}

\def\a{\alpha}
\def\b{\beta}

\def\d{\delta}

\def\g{\gamma}

\def\j{\psi}
\def\k{\kappa}
\def\l{\lambda}

\def\o{\omega}

\def\q{\theta}
\def\r{\rho}
\def\s{\sigma}
\def\t{\tau}

\def\D{\Delta}

\def\L{\Lambda}

\def\S{\Sigma}

\def\tr{{\rm tr}}

\def\ri{{\rm i}}

\newcommand{\ve}{\varepsilon}                            

\newcommand{\pa}{\partial}                           
\newcommand{\hf}{\frac12}

\newcommand{\be}{\begin{equation}}
\newcommand{\ee}{\end{equation}}
\newcommand{\bea}{\begin{eqnarray}}
\newcommand{\eea}{\end{eqnarray}}
\newcommand{\non}{\nonumber}
\newcommand{\ba}{\begin{array}}
\newcommand{\ea}{\end{array}}

\newcommand{\1}{{\underline{1}}}

\newcommand{\dsR}{{\mathbb R}}

\newcommand{\bm}[1]{\mbox{\boldmath$#1$}}

\def\double #1{#1{\hbox{\kern-2pt $#1$}}}

\newcommand{\bsubeq}{\begin{subequations}}
\newcommand{\esubeq}{\end{subequations}}

\newcommand{\ul}{\underline}
\newcommand{\eps}{{\ve}}

\newcommand{\dalpha}{{\dot{\alpha}}}

\newcommand{\rd}{\mathrm d}
\numberwithin{equation}{section}

\newcommand{\RM}{R(M)}
\newcommand{\RD}{R(\mathbb D)}
\newcommand{\RN}{R(N)}
\newcommand{\RJ}{R(\cJ)}
\newcommand{\RS}{R(S)}
\newcommand{\RK}{R(K)}

\begin{document}

\begin{titlepage}
\begin{flushright}
Nikhef-2013-014\\
May, 2013\\
\end{flushright}
\vspace{5mm}

\begin{center}
{\Large \bf 
Conformal supergravity in three dimensions: \\
New off-shell formulation}
\\ 
\end{center}

\begin{center}

{\bf
Daniel Butter${}^{a}$, Sergei M. Kuzenko${}^{b}$, Joseph Novak${}^{b}$, \\
and
Gabriele Tartaglino-Mazzucchelli${}^{b}$
} \\
\vspace{5mm}

\footnotesize{
${}^{a}${\it Nikhef Theory Group \\
Science Park 105, 1098 XG Amsterdam, The Netherlands}}
~\\
\texttt{dbutter@nikhef.nl}\\
\vspace{2mm}

\footnotesize{
${}^{b}${\it School of Physics M013, The University of Western Australia\\
35 Stirling Highway, Crawley W.A. 6009, Australia}}  
~\\
\texttt{joseph.novak,\,gabriele.tartaglino-mazzucchelli@uwa.edu.au}\\
\vspace{2mm}

\end{center}

\begin{abstract}
\baselineskip=14pt
We propose a new off-shell formulation for $\cN$-extended conformal supergravity 
in three spacetime dimensions. Our construction is based on the gauging of the 
$\cN$-extended superconformal algebra in superspace. Covariant constraints are 
imposed such that the algebra of covariant derivatives is given in terms of a single 
curvature superfield which turns out to be the super Cotton tensor. 
An immediate corollary of this construction is that
the curved superspace is conformally flat
if and only if  the super Cotton tensor vanishes.
Upon degauging of certain local symmetries, 
our formulation is shown to reduce
to the conventional one 
with the local structure group $\rm SL(2 , \dsR) \times SO(\cN)$.
\end{abstract}

\vfill

\vfill
\end{titlepage}

\newpage
\renewcommand{\thefootnote}{\arabic{footnote}}
\setcounter{footnote}{0}

\tableofcontents



\section{Introduction}

Inspired by the construction of topologically massive $\cN=1$ supergravity in three dimensions 
$(3D)$ \cite{DK,Deser},  conformal supergravity theories in $3D$
were formulated as supersymmetric Chern-Simons theories for  $\cN=1$ \cite{vanN85}, 
$\cN=2$ \cite{RvanN86},\footnote{In the $\cN=1$ case, 
the  superconformal tensor calculus was independently developed in \cite{Uematsu}.
Early superspace approaches to $\cN=1$ and $\cN=2$ supergravity theories 
were given in \cite{BG,GGRS,ZP,ZP89}.} 
and finally for arbitrary $\cN$ \cite{LR89,NG}.\footnote{We are grateful to Jim Gates for 
bringing Ref. \cite{NG} to our attention. }
The constructions in  \cite{vanN85,RvanN86,LR89,NG} are based on the gauging 
of the $\cN$-extended
superconformal algebra in ordinary spacetime.  
The important point is that the formulation for extended conformal supergravity given in \cite{LR89,NG} is on-shell for $\cN>2$.
This means that alternative approaches are required if one is interested in deriving off-shell actions for extended conformal supergravity, especially in the presence of matter. 

In 1995 Howe {\it et al.} \cite{HIPT} proposed a curved superspace geometry with local 
structure group  $\rm SL(2 , \dsR) \times SO(\cN)$,
which is suitable for the  description of off-shell  $\cN$-extended conformal supergravity in three dimensions.\footnote{This construction is a natural generalization of Howe's 
superspace formulation  
for $4D$  $\cN$-extended conformal supergravity in four dimensions \cite{Howe}.}
Specifically, Ref. \cite{HIPT} postulated the superspace constraints, 
determined all components of the superspace torsion of dimension-1, 
and identified the component $\cN$-extended Weyl supermultiplet.
At the same time, some crucial  elements of the formalism 
(including the explicit structure of super-Weyl transformations and the solution of 
the dimension-3/2 and dimension-2 Bianchi identities)
did not appear
in \cite{HIPT}. The geometry of $\cN$-extended conformal supergravity has recently 
been fully developed in \cite{KLT-M11}\footnote{The special cases of $\cN=8$ and $\cN=16$ conformal supergravity theories 
were independently  worked out in 
\cite{Howe:2004ib,CGN} and \cite{GH} respectively.} 
 and applied  to the construction 
 of general supergravity-matter couplings in the cases $\cN \leq 4$
(the simplest extended case $\cN=2$ was studied in more detail in \cite{KT-M11}).

It turns out that the problem of constructing off-shell superspace actions 
for pure extended conformal supergravity theories is rather nontrivial.  
The action for $\cN=1$ conformal supergravity can readily be derived in terms of the 
superfield connection as a superspace integral \cite{GGRS,ZP,ZP89}
(although the  results in  \cite{GGRS,ZP,ZP89} are incomplete, and the conformal supergravity 
action has only recently been given in \cite{KT-M12}).
However, such a construction becomes impossible starting from $\cN=2$.\footnote{If 
a prepotential formulation is available, the conformal supergravity action may be written 
as a superspace integral in terms of the prepotentials.}
As discussed in \cite{KT-M12}, 
this is because (i)  the spinor and vector sectors of the superfield connection 
have  positive dimension equal to $1/2$ and 1
respectively; 
and (ii) the dimension of the full superspace measure is  $(\cN- 3)$. 
This implies that it is not possible to construct 
contributions to the action 
that are cubic in the superfield connection for $\cN\geq 2$. 

Nevertheless, it was argued by two of us \cite{KT-M12} that 
$\cN$-extended conformal supergravity 
can be realized in terms of the off-shell 
Weyl supermultiplet \cite{HIPT} and the associated curved superspace geometry 
\cite{HIPT,KLT-M11}. 
Such a realization was explicitly worked out in \cite{KT-M12} for the case $\cN=1$, 
and a general method  of constructing conformal supergravity actions for $\cN>1$ was outlined. 
It should be pointed out that the approach  of \cite{KT-M12} 
is a generalization of the superform formulation for the linear multiplet in 
four-dimensional $\cN=2$ conformal supergravity given in \cite{BKN}. 
Both works  \cite{KT-M12,BKN} make use of  the superform approach for
the construction of supersymmetric invariants \cite{Hasler,Ectoplasm,GGKS}, also known as 
 the ectoplasm formalism \cite{Ectoplasm,GGKS}.

It is worth recalling the method sketched in \cite{KT-M12}. 
Let $\cD_A =(\cD_a, \cD_\a^I)$ be the superspace covariant derivatives, with $I=1,\dots,  \cN$, 
which describe the off-shell $\cN$-extended Weyl supermultiplet \cite{HIPT,KLT-M11}. 
Following the conventions of \cite{KLT-M11}, one should start with  a
two-parameter deformation of the vector covariant derivative
\bea
\cD_{\a\b} ~\to ~ {\mathfrak D}_{\a\b} = \cD_{\a\b} +\l \cS M_{\a\b} + \r C_{\a\b}{}^{KL} N_{KL}~,
\label{1.1}
\eea
where $\l$ and $\r$ are real parameters, and $\cS$ and $ C_{\a\b}{}^{KL} $ are certain dimension-1
torsion tensors.  
The deformed covariant derivatives ${\mathfrak D}_A =({\mathfrak D}_a,  {\mathfrak D}_\a^I)
:= ({\mathfrak D}_a, \cD_\a^I)$ 
obey the algebra
\bea
{[} {\mathfrak D}_{{A}}, {\mathfrak D}_{{B}}\}&=& -{\bm T}_{{A}{B}}{}^{{C}} {\mathfrak D}_{{C}}
-\hf {\bm R}_{{A}{B}}{}^{cd}M_{cd}
-\hf {\bm R}_{AB}{}^{KL}N_{KL}~,
\eea
with  ${\bm T}_{AB}{}^C$ the torsion,  ${\bm R}_{AB}{}^{cd}$ the Lorentz curvature and
 ${\bm R}_{AB}{}^{KL}$ the SO($\cN$) curvature.\footnote{Our conventions for sign of
the torsion, curvatures and connections differ from those in \cite{KLT-M11, KT-M12}.}
As a next stage, one has to consider the superform equation
\bea
\rd {\bm \S}= \hf {\bm R}^{ab} \wedge {\bm R}_{ab} + \frac{\k}{2}  {\bm R}^{IJ} \wedge {\bm R}_{IJ} ~,
\eea
with $\k$  a real parameter, and look for two solutions ${\bm \S}_{T}$ and ${\bm \S}_{\rm CS}$.
Here ${\bm \S}_{T}$ is a three-form constructed in terms of the torsion and curvature 
tensors and their covariant derivatives, while   ${\bm \S}_{\rm CS}$ 
is a standard Chern-Simons three-form. 
Now, the three-form ${\bm \S} := {\bm \S}_{T} - {\bm \S}_{\rm CS}$ has the following properties
(i) ${\bm \S} $  is closed;  and (ii) ${\bm \S} $ is 
a polynomial in two variables $\l$ and $\r$. By differentiating $\bm \S$ with respect to $\l$ and $\r$, 
we will generate
 a number of closed three-forms. Finally, one has to look for a linear combination
$\mathfrak J$ of these closed three-forms, which is {\it super-Weyl  invariant}  modulo exact contributions. 
The parameter $\k$ is expected to be fixed by this requirement. 
It is also expected that ${\mathfrak J}$ is independent 
of $\lambda$ and $\rho$, due to its uniqueness. 
The closed three-form $\mathfrak J$ generates the action for $\cN$-extended conformal supergravity. 

In the previous paper \cite{KT-M12},
 the above method was applied only in the case $\cN=1$. In this and only this case, 
there is no $\r$-deformation. In spite of this simplification, the calculation of $\mathfrak J$
was rather long and tedious.  Two of us (SMK and GT-M) have tried to apply the same method
in order to construct the action for $\cN=2$ conformal supergravity. 
The  computation required turned out to be extremely involved.   
This essentially means that the curved superspace geometry of
\cite{HIPT,KLT-M11} is not well adapted for the construction of conformal supergravity actions, 
and  we should look for an alternative formulation for $\cN$-extended conformal supergravity. 

In the present paper, we propose a new off-shell formulation 
for $\cN$-extended conformal supergravity in three dimensions. 
It is inspired by the recently developed formulations for $\cN=1$ \cite{ButterN=1}
and $\cN=2$ \cite{ButterN=2} conformal supergravities in four dimensions. 
These formulations are obtained by gauging the superconformal algebra in superspace. 
Conceptually such a gauging is similar to the superconformal tensor calculus 
in the component setting (see, e.g.,  \cite{FradTsey,FVP} for reviews).    
The crucial new point of the $4D$ superspace approaches in  \cite{ButterN=1, ButterN=2}
is that covariant constraints are  imposed such that the algebra of covariant derivatives is given in terms of a single curvature superfield which coincides with the super Weyl tensor. 
This turns out to lead to dramatic computational simplifications. 

This paper is organized as follows.
Section \ref{setup} describes the geometric setup of $\cN$-extended conformal superspace in 
three dimensions. We present the superconformal algebra and the procedure in which it is gauged within superspace. In 
section \ref{confGrav} we provide a warm-up construction. As a straightforward extension of the gauging procedure for the bosonic case, 
we describe the geometry of conformal gravity for $D \geq 3$. In section \ref{confSUGRA} we show how to constrain the geometry of 
section \ref{setup} to describe $\cN$-extended conformal supergravity, thus providing a new off-shell formulation in superspace. 
Section \ref{Degauging} is dedicated to showing how the conventional superspace formulation of \cite{HIPT, KLT-M11} may be viewed as 
a degauged version of the $\cN$-extended conformal superspace. Finally, section \ref{conclusion} concludes the paper by discussing the 
newly obtained results. 

We have included a couple of technical appendices. In appendix \ref{NC} we include a summary of 
our notation and conventions. Appendix \ref{VM} shows how to couple an Abelian $\cN$-extended vector multiplet to 
conformal supergravity within our superspace formulation.


\section{Setup for $\cN$-extended conformal superspace} \label{setup}

In this section we 
present a geometric setup for 
$\cN$-extended conformal superspace in three dimensions ($3D$), 
which arises from gauging the 
$\cN$-extended superconformal algebra. 
We begin our discussion by giving the $\cN$-extended superconformal algebra in our 
notation and conventions. We then present the gauging procedure for the construction of 
conformal superspace, 
which parallels the previous work in four dimensions \cite{ButterN=1, ButterN=2}.


\subsection{$\cN$-extended superconformal algebra in three dimensions}

The bosonic part of the $3D$ $\cN$-extended superconformal algebra \cite{Nahm},
 ${\mathfrak{osp}}(\cN|4, {\mathbb R})$,
contains the translation ($P_a$), Lorentz ($M_{ab}$), special conformal ($K_a$), 
dilatation ($\mathbb D$) and $\rm SO(\cN)$ ($N_{KL}$) generators, where 
$K,L=1,\dots, \cN$. 
Their algebra is
\begin{subequations} \label{SCA}
\begin{gather}
[M_{ab} , M_{cd}] = 2 \eta_{c[a} M_{b] d} - 2 \eta_{d [a} M_{b] c} \ ,\label{SCA.a} \\
[M_{ab} , P_c ] = 2 \eta_{c [a} P_{b]} \ , \quad [\mathbb D, P_a] = P_a \ , \\
[M_{ab} , K_c] = 2 \eta_{c[a} K_{b]} \ , \quad  [\mathbb D, K_a] = - K_a \ , \\
[K_a , P_b] = 2 \eta_{ab} \mathbb D + 2 M_{ab} \ , \\
[N_{KL} , N^{IJ}] = 2 \d^I_{[K} N_{L]}{}^J - 2 \d^J_{[K} N_{L]}{}^I \ , \label{SCA.e}
\end{gather}
where all other commutators vanish. The extension
to the superconformal case is achieved by extending the translation 
generator to $P_A = (P_a , Q_\a^I)$ and the special conformal generator to
$K_A = (K_a , S_\a^I)$,
where $Q_\a^I$ and
$S_\a^I$
are $3D$ spinors with respect to the index $\a$ and $\rm SO(\cN)$ vectors with respect to the index $I$ (see 
appendix \ref{NC}).\footnote{In line with usual nomenclature we
refer to $S_\a^I$ as the $S$-supersymmetry generator and $K_a$ as the special conformal 
boost. We will also frequently refer to the full set $K_A = (K_a, S_\alpha^I)$ as the
special conformal generator where there is little ambiguity.} The fermionic generator $Q_\a^I$ obeys the
algebra
\begin{gather}
\{ Q_\a^I \ , Q_\b^J \} = 2 \ri \d^{IJ} (\g^c)_{\a\b} P_c = 2 \ri \d^{IJ} P_{\a\b} \ , \quad [Q_\a^I , P_b ] = 0 \ , \\
[M_{\a\b} , Q_\g^I] = \eps_{\g(\a} Q_{\b)}^I \ , \quad [\mathbb D, Q_\a^I] = \hf Q_\a^I \ , \quad [N_{KL} , Q_\a^I] = 2 \d^I_{[K} Q_{\a L]} \ ,
\end{gather}
while the generator $S_\a^I$ obeys the
algebra
\begin{gather}
\{ S_\a^I , S_\b^J \} = 2 \ri \d^{IJ} (\g^c)_{\a\b} K_c \ , \quad [S_\a^I , K_b] = 0 \ , \\
[M_{\a\b} , S_\g^I] = \eps_{\g(\a} S_{\b)}^I \ , \quad [\mathbb D, S_\a^I] = - \hf S_\a^I \ , \quad [N_{KL} , S_\a^I] = 2 \d^I_{[K} S_{\a L]} \ .
\end{gather}
Finally, the 
remainder of the algebra of $K_A$ with $P_A$ is given by
\begin{gather}
[K_a , Q_\a^I ] = - \ri (\g_a)_\a{}^\b S_\b^I \ , \quad [S_\a^I , P_a] = \ri (\g_a)_\a{}^{\b} Q_{\b}^I \ , \\
\{ S_\a^I , Q_\b^J \} = 2 \eps_{\a\b} \d^{I J} \mathbb D - 2 \d^{I J} M_{\a\b} - 2 \eps_{\a\b} N^{IJ} \ .
\end{gather}
\end{subequations}
For a matrix realization of the $\cN$-extended superconformal algebra, 
see e.g. \cite{KPT-MvU}.

The superconformal algebra\footnote{Up to conventions and for the $\cN = 1$ case the algebra \eqref{SCA} agrees with \cite{vanN85}.}
must obey the Jacobi identities.
If we denote the generators of the algebra by $X_{\tilde{a}}$ then the Jacobi identities may be written as
\begin{align} [X_{\tilde{a}}, [X_{\tilde{b}}, X_{\tilde{c}} \} \} + (-1)^{\eps_{\tilde{a}} (\eps_{\tilde{b}} + \eps_{\tilde{c}})} [X_{\tilde{b}}, [X_{\tilde{c}}, X_{\tilde{a}} \} \}
+ (-1)^{\eps_{\tilde{c}} (\eps_{\tilde{a}} + \eps_{\tilde{b}})} [X_{\tilde{c}} , [ X_{\tilde{a}} , X_{\tilde{b}} \} \} = 0 \ ,
\end{align}
where $\eps_{\tilde{a}} = \eps(X_{\tilde{a}})$ is the Grassmann parity of $X_{\tilde{a}}$. If we further denote the algebra \eqref{SCA} by\footnote{When summing over the pairs 
of antisymmetric Lorentz and $\rm SO(\cN)$ indices there is a factor of a half which is suppressed here.}
\be
[X_{\tilde{a}}, X_{\tilde{b}}] = -f_{\tilde{a}\tilde{b}}{}^{\tilde{c}} X_{\tilde{c}} \ ,
\ee
where $f_{\tilde{a}\tilde{b}}{}^{\tilde{c}} = - (-1)^{\eps_{\tilde{a}} \eps_{\tilde{b}}} f_{\tilde{b}\tilde{a}}{}^{\tilde{c}}$ are the structure constants, then 
we may equivalently write the Jacobi identities as
\be f_{[\tilde{a}\tilde{b}}{}^{\tilde{d}} f_{|\tilde{d}| \tilde{c} \} }{}^{\tilde{e}} = 0 \ .
\ee

The remainder of our notation and conventions follow closely those of \cite{KLT-M11} and are summarized in appendix \ref{NC}.


\subsection{Gauging the superconformal algebra} \label{Gauging}

To perform our gauging procedure, we begin with a curved $3D$ $\cN$-extended superspace
 $\cM^{3|2 \cN}$ parametrized by
local bosonic $(x)$ and fermionic coordinates $(\theta_I)$:
\be z^M = (x^m, \ \q^\mu_I) \ ,
\ee
where $m = 0, 1, 2$, $\mu = 1, 2$ and $I = 1, \cdots , \cN$. In order to describe supergravity it is necessary to 
have built into the theory a vielbein and appropriate connections. However the gauging of the superconformal algebra is made  
non-trivial due to the fact that the graded commutator of
$K_A$ with $P_A$ contains generators other
than $P_A$.
This requires some of the connections to transform under $K_A$ into the vielbein.
To perform the gauging we will  follow closely the approach
given in \cite{ButterN=2}.

In order to gauge the superconformal algebra\footnote{Gauging spacetime symmetries differs from the usual approach employed for internal symmetry groups. 
The superspace approach here follows closely the one described in \cite{ButterN=2}.} it is useful to denote by $X_{\underline{a}}$ 
the subset of the generators which do not contain the $P_A$ generators. 
The superconformal algebra may be written as
\bsubeq
\begin{align}
[X_{\underline{a}} , X_{\underline{b}} \} &= -f_{\underline{a} \underline{b}}{}^{\underline{c}} X_{\underline{c}} \ , \\
[X_{\underline{a}} , P_B \} &= -f_{\underline{a} B}{}^{\underline{c}} X_{\underline{c}}
	- f_{\underline{a} B}{}^C P_C := -f_{\underline{a} P_B}{}^{\tilde{c}} X_{\tilde{c}} \ , \\
[P_A , P_B \} &= -f_{A B}{}^{C} P_C := - f_{P_A P_B}{}^{P_C} P_C  \ ,
\end{align}
\esubeq
where $f_{AB}{}^C$ contains only the constant torsion tensor
$f_{\a}^I{}_\b^J{}^c = T_\a^I{}_\b^J{}^c = -2 \ri \d^{IJ} (\g^c)_{\a\b}$.
It is seen that the generators  $X_{\underline{a}}$ form a superalgebra. 
The gauge group associated with the superalgebra will be denoted $\cH$. 

In order to gauge the algebra \eqref{SCA} we associate with each generator $X_{\underline{a}}$ a connection one-form 
$\omega^{\underline{a}} = \rd z^M \omega_M{}^{\underline{a}}$ and with $P_A$ the vielbein 
$E^A = \rd z^M E_M{}^A$. 
Their gauge transformations are postulated to be
\begin{subequations} \label{VGCTR}
\begin{align}
\d_{\cH} E^A &= E^B \L^{\underline{c}} f_{\underline{c} B}{}^A \ , \\
\d_{\cH} \omega^{\underline{a}} &= \rd \L^{\underline{a}} 
+ E^B \L^{\underline{c}} f_{\underline{c} B}{}^{\underline{a}} 
+ \omega^{\underline{b}} \L^{\underline{c}} f_{\underline{c} \underline{b}}{}^{\underline{a}} \ ,
\end{align}
\end{subequations}
with
$\L^{\underline{a}}$ the gauge parameters. 
A superfield $\Phi$ is said to be {\it covariant} 
if it transforms under $\cH$ with no derivative of the parameter $\L^{\underline{a}}$
\be \d_{\cH} \Phi = \L \Phi := \L^{\underline{a}} X_{\underline{a}} \Phi \ .
\ee
If $\Phi$ transforms in some tensor representation of
$\cH$ we have matrix 
realizations
\be 
M_{ab} \Phi = m_{ab} \Phi \ , \quad 
N_{IJ} \Phi = n_{IJ} \Phi \ , \quad \mathbb D \Phi = \D \Phi \ , \quad 
\ee
where $\D$ is a real number corresponding to the conformal dimension,
 and $m_{ab}$  and $n_{IJ}$ are the  Lorentz and isospin matrices 
 associated with $\Phi$.\footnote{These matrices obey the commutation relations
  $[m_{ab} , m_{cd}] = - 2 \eta_{c[a} m_{b] d} + 2 \eta_{d [a} m_{b] c} $ and 
  $[n_{KL} , n^{IJ}] = -2 \d^I_{[K} n_{L]}{}^J + 2 \d^J_{[K} n_{L]}{}^I $, compare 
  with \eqref{SCA.a} and \eqref{SCA.e}.
  }
The final generators $K_A = (K_a, S_\alpha^I)$ are used to define conformal {\it primary} superfields:
\be 
K_A \Phi = 0 \ .
\ee
From the algebra, we note that if a superfield is annihilated by $S$-supersymmetry,
then it is necessarily primary.

It is obvious that $\pa_M \Phi$ is not itself covariant. We are led to introduce
the {\it covariant derivative}
\be
\nabla = \rd - \omega^{\underline{a}} X_{\underline{a}} \ , \quad \nabla = E^A \nabla_A \ .
\ee
Its transformation is found to be
\be
\d_{\cH} (\nabla_A \Phi) = (-1)^{\eps_A \eps_{\underline{b}}} \L^{\underline{b}} \nabla_A X_{\underline{b}} \Phi - \L^{\underline{b}} f_{\underline{b} A}{}^C \nabla_C \Phi 
- \L^{\underline{b}} f_{\underline{b} A}{}^{\underline{c}} X_{\underline{c}} \Phi \ ,
\ee
with no derivatives on the gauge parameter $\L^{\underline{a}}$. 
Rewriting this as 
$\d_{\cH} (\nabla_A \Phi) = \L^{\underline{b}} X_{\underline{b}} \nabla_A \Phi$, 
we immediately find the operator relation
\be
[ X_{\underline{b}} , \nabla_A \} = -f_{\underline{b} A}{}^C \nabla_C
	- f_{\underline{b} A}{}^{\underline{c}} X_{\underline{c}} \ .
\ee
The curvature and torsion tensors appear in the commutator of two covariant derivatives,
\be
[ \nabla_A , \nabla_B \} = - T_{AB}{}^C \nabla_C - R_{AB}{}^{\underline{c}} X_{\underline{c}} \ .
\ee
The explicit expressions for these tensors
are most compactly given in terms of  two-forms
\begin{subequations}
\begin{align}
T^A &:= \hf E^C \wedge E^B T_{BC}{}^A = \rd E^A - E^C \wedge \omega^{\underline{b}} \,f_{\underline{b} C}{}^A \ , \\
R^{\underline{a}} &:= \hf E^C \wedge E^B R_{BC}{}^{\underline{a}} = \rd \omega^{\underline{a}}
	- E^C \wedge \omega^{\underline{b}} \, f_{\underline{b} C}{}^{\underline{a}}
	- \hf \omega^{\underline{c}} \wedge \omega^{\underline{b}} \,
		f_{\underline{b} \underline{c}}{}^{\underline{a}} \ .
\end{align}
\end{subequations}
Using the definition of curvature and torsion 
together with the vielbein and connection transformation rules \eqref{VGCTR} we find
\begin{subequations}
\begin{align}
\d_\cH T^A &= T^C \L^{\underline{b}} f_{\underline{b} C}{}^A
	- E^C \wedge E^B \L^{\underline{a}} f_{\underline{a} B}{}^{\underline{f}} f_{\underline{f} C}{}^A \ , \\
\d_{\cH} R^{\underline{a}} &=
	R^{\underline{c}} \L^{\underline{b}} f_{\underline{b} \underline{c}}{}^{\underline{a}}
	+ T^C \L^{\underline{b}} f_{\underline{b} C}{}^{\underline{a}}
	- E^D \wedge E^C \L^{\underline{b}}
		f_{\underline{b} C}{}^{\underline{f}} f_{\underline{f} D}{}^{\underline{a}} \ .
\end{align}
\end{subequations}
Writing the transformation rules as 
$\d_\cH T^A = \L^{\underline{a}} X_{\underline{a}} T^A$, 
$\d_\cH R^A  = \L^{\underline{a}} X_{\underline{a}} R^A$ and 
$\d_\cH E^A = \L^{\underline{b}} X_{\underline{b}} E^A$ 
leads to the action of $X_{\underline{a}}$ on the torsion and curvature:
\begin{subequations}
\begin{align}
X_{\underline{a}} T_{BC}{}^D =&
	- (-1)^{\eps_{\underline{a}} (\eps_{B}+\eps_{C})}  T_{BC}{}^F f_{F \underline{a}}{}^D
	- 2 f_{\underline{a} [B}{}^F T_{|F| C\}}{}^D
	- 2 f_{\underline{a} [B}{}^{\underline{f}} f_{|\underline{f}| C\}}{}^D \ , \\
X_{\underline{a}} R_{BC}{}^{\underline{d}} =&
	- (-1)^{\eps_{\underline{a}} (\eps_{B}+\eps_{C})} \Big(T_{BC}{}^F f_{F \underline{a}}{}^{\underline{d}}
	+ R_{BC}{}^{\underline{f}} f_{\underline{f}\underline{a} }{}^{\underline{d}}\Big)
	- 2 f_{\underline{a} [B}{}^F R_{|F| C \}}{}^{\underline{d}} \non\\
	&- 2 f_{\underline{a} [B}{}^{\underline{f}} f_{|\underline{f}| C \}}{}^{\underline{d}} \ .
\end{align}
\end{subequations}

One can show the above results are the necessary conditions for the Jacobi identity involving two $\nabla$'s
\be 0 = [ X_{\underline{a}} , [ \nabla_B , \nabla_C \} \} + {\rm cycles}
\ee
to be identically satisfied. The Bianchi identities
\be 0 = [ \nabla_A , [ \nabla_B , \nabla_C \} \} + {\rm cycles} 
\ee
can also be shown to be satisfied identically.
Therefore, we have a consistent algebraic structure\footnote{The reason why the sign of the structure constants 
was chosen was so that in the flat limit the torsion becomes the usual structure constant for the $[ P_A , P_B \}$ (anti-)commutator.}
\begin{subequations}\label{2.20}
\begin{align}
[X_{\underline{a}} , X_{\underline{b}} \} &
= -f_{\underline{a} \underline{b}}{}^{\underline{c}} X_{\underline{c}} \ , \\
[X_{\underline{a}} , \nabla_B \} &= - f_{\underline{a} B}{}^C \nabla_C -f_{\underline{a} B}{}^{\underline{c}} X_{\underline{c}} \ , \label{eq:XwithNabla} \\
[\nabla_A , \nabla_B \} &= -T_{AB}{}^C \nabla_C - R_{AB}{}^{\underline{c}} X_{\underline{c}} \ ,
\end{align}
\end{subequations}
which satisfies all the Jacobi identities.
In the flat space limit the curvature vanishes and the torsion becomes the usual constant torsion, 
so that the algebra \eqref{2.20} exactly matches the 
superconformal algebra that we started with, in which $P_A$ is replaced with $\nabla_A$. 
The curved case requires a deformation via the introduction of torsion and curvature. 
The superconformal algebra is then said to be ``gauged'' in this sense.

The full set of operators $(\nabla_A, X_{\ul a})$ generates the conformal supergravity gauge group $\cG$.
The form of the covariant derivative suggests that we should extend the usual
diffeomorphisms $\delta_{\textrm{gct}}$ into \emph{covariant diffeomorphisms}\footnote{These transformations are also known as 
covariant general coordinate transformations. Their use is standard, see {\rm e.g.} \cite{FVP}.}
\begin{align}
\delta_{\rm cgct}(\xi^A) := \delta_{\textrm{gct}} (\xi^A E_A{}^M) - \delta_{\cH}(\xi^A \omega_A{}^{\ul a})~,
\end{align}
where $\delta_{\textrm{gct}}(\xi^M)$
acts on scalars under diffeomorphisms as
\begin{align} \delta_{\textrm{gct}} \Phi = \xi^M \partial_M \Phi \ .
\end{align}
The full conformal supergravity gauge group $\cG$ is then generated by
\begin{align}
\cK = \xi^C \nabla_C + \L^{\ul a} X_{\ul a}~.
\end{align}
If a superfield $\Phi$ is a scalar under diffeomorphisms and covariant under the group $\cH$,
then its transformation under the full supergravity gauge group $\cG$ is
\begin{align}
\d_\cG \Phi = \cK \Phi = \xi^C \nabla_C \Phi + \L^{\ul a} X_{\ul a} \Phi~.
\end{align}
It is a straightforward exercise to show that the vielbein and connection one-forms
transform as
\begin{subequations}\label{eq:deltaConn}
\begin{align}
\delta_\cG E^A &= \rd \xi^A + E^B \L^{\ul c} f_{\ul c B}{}^A
	+ \omega^{\ul b} \xi^{C} f_{C \ul b}{}^{A}
	+ E^B \xi^{C} T_{C B}{}^A~, \\
\delta_\cG \omega^{\ul a} &= \rd \L^{\ul a}
	+ \omega^{\ul b} \L^{\ul c} f_{\ul c \ul b}{}^{\ul a}
	+ \omega^{\ul b} \xi^{C} f_{C \ul b}{}^{\ul a}
	+ E^B \L^{\ul c} f_{\ul c B}{}^{\ul a}
	+ E^B \xi^{C} R_{C B}{}^{\ul a}~.
\end{align}
\end{subequations}
${}$From this definition, one can check that the covariant derivative transforms as
\begin{align}\label{TransCD}
\delta_\cG \nabla_A = [\cK,\nabla_A]
\end{align}
provided we interpret\footnote{One must take care in applying the formulae \eqref{TransCD} and \eqref{eq:nablaParams}. 
Observe that we can have $\L^{\ul b} = 0$ but $\nabla_A \L^{\ul b} \neq 0$ if
either $\xi^D f_{D \ul c}{}^{\ul b}$ or $\L^{\ul d} f_{\ul d \ul c}{}^{\ul b}$ is
non-vanishing.}
\begin{subequations}\label{eq:nablaParams}
\begin{align}
\nabla_A \xi^B &:= E_A \xi^B + \omega_A{}^{\ul c} \xi^D f_{D \ul c}{}^B~, \\
\nabla_A \L^{\ul b} &:= E_A \L^{\ul b}
	+ \omega_A{}^{\ul c} \xi^D f_{D \ul c}{}^{\ul b}
	+ \omega_A{}^{\ul c} \L^{\ul d} f_{\ul d \ul c}{}^{\ul b}
	~.
\end{align}
\end{subequations}

We can summarize the superspace geometry of conformal supergravity
as follows. The covariant derivatives have the form\footnote{Note that the complex conjugation rule \eqref{CDconjR} induces a natural 
reality condition on the vielbein and the connections.}
\be
\nabla_A = E_A - \o_A{}^{\underline b} X_{\underline b} 
= E_A - \hf \Omega_A{}^{ab} M_{ab} - \hf \Phi_A{}^{PQ} N_{PQ} - B_A \mathbb D - \mathfrak{F}_A{}^B K_B \ .
\ee
The action of the generators on the covariant derivatives, eq. \eqref{eq:XwithNabla},
resembles that for the $P_A$ generators given in \eqref{SCA}.
The supergravity gauge group is generated by local transformations of the form
\eqref{TransCD}
where
\bea
\cK &=& \xi^C \nabla_C + \hf \L^{cd} M_{cd} + \hf \L^{PQ} N_{PQ} + \s \mathbb D 
+ \L^A K_A  
\eea
and the gauge parameters satisfy natural reality conditions.
The covariant derivatives satisfy the (anti-)commutation relations
\begin{align}
[ \nabla_A , \nabla_B \}
	&= -T_{AB}{}^C \nabla_C
	- \frac{1}{2} \RM_{AB}{}^{cd} M_{cd}
	- \frac{1}{2} \RN_{AB}{}^{PQ} N_{PQ}
	\non \\ & \quad
	- \RD_{AB} \mathbb D
	- \RS_{AB}{}^\g_K S_\g^K
	- \RK_{AB}{}^c K_c~,
\end{align}
where the torsion and curvature tensors are given by\footnote{Since $\rm SO(\cN)$ vector indices are raised and lowered using the Kronecker delta, there is no need to distinguish 
between upper and lower $\rm SO(\cN)$ vector indices.}
\begin{subequations}
\bea
T^a &=& \rd E^a + E^a \wedge B + E^b \wedge \Omega_b{}^a \ , \\
T^\a_I &=& \rd E^\a_I + \hf E^\b_I \wedge \Omega^{c} (\g_c)_\b{}^\a + \hf E^\a_I \wedge B + E^{\a J} \wedge \Phi_{JI} + \ri \, E^c \wedge \mathfrak{F}^\beta_I (\gamma_c)_\beta{}^\alpha \ ,~~~~~~~~~~~ \\
\RD &=& \rd B + 2 E^a \wedge \mathfrak{F}_a - 2 E^\a_I \wedge \mathfrak{F}_\a^I \ , \\
\RM^{ab} &=& \rd \Omega^{ab} + \Omega^{ac} \wedge \Omega_c{}^b - 4 E^{[a} \wedge \mathfrak{F}^{b]} - 2 E^\a_I \wedge \mathfrak{F}^{\b I} (\g_c)_{\a\b} \eps^{cab} \ , \\
\RN^{IJ} &=& \rd \Phi^{IJ} + \Phi^{I K} \wedge \Phi_{K}{}_J - 4 E^{\a [I} \wedge \mathfrak{F}_{\a}{}^{J]} \ , \\
\RK^a &=& \rd \mathfrak{F}^a - \mathfrak{F}^a \wedge B + \mathfrak{F}^b \wedge \Omega_b{}^a + \ri \mathfrak{F}^\a_I \wedge \mathfrak{F}^{\b I} (\g^a)_{\a\b} \ , \\
\RS^\a_I &=& \rd \mathfrak{F}^\a_I - \ri E^\b_I \wedge \mathfrak{F}^a (\g_a)_\b{}^\a - \hf \mathfrak{F}^\a_I \wedge B + \hf \mathfrak{F}^\b_I \wedge \Omega^{c} (\g_c)_\b{}^\a + \mathfrak{F}^{\a J} \wedge \Phi_{JI}     \ .
\eea
\end{subequations}


\section{Conformal gravity in $D \geq 3$ dimensions} \label{confGrav} 

Before we turn to $3D$ conformal supergravity 
we will first discuss conformal gravity in $D \geq 3$ dimensions.\footnote{Conformal gravity has been discussed elsewhere in many places, \rm e.g. \cite{FVP}. Here
we review conformal gravity emphasizing some points relevant to our paper.
The important feature is that the algebra of covariant derivatives may be constructed entirely in terms of a 
primary superfield.} 
To do so we note that 
the bosonic part of the superconformal algebra \eqref{SCA} without the $\rm SO(\cN)$ generator can be straightforwardly extended to $D$ 
dimensions. The algebra is
\begin{subequations}
\begin{gather}
[M_{ab} , M_{cd}] = 2 \eta_{c[a} M_{b] d} - 2 \eta_{d [a} M_{b] c} \ , \\
[M_{ab} , P_c ] = 2 \eta_{c [a} P_{b]} \ , \quad [\mathbb D, P_a] = P_a \ , \\
[M_{ab} , K_c] = 2 \eta_{c[a} K_{b]} \ , \quad  [\mathbb D, K_a] = - K_a \ , \\
[K_a , P_b] = 2 \eta_{ab} \mathbb D + 2 M_{ab} \ ,
\end{gather}
\end{subequations}
where all other commutators vanish and $\eta_{ab}$ is the the $D$-dimensional Minkowski metric. It is clear that the gauging procedure of 
section \ref{Gauging} may be straightforwardly extended to conformal gravity in $D$ dimensions, while the restriction to a bosonic manifold 
is trivial.
 
The covariant derivatives
have the form
\be
\nabla_a =  e_a - \hf \omega_a{}^{bc} M_{bc} - b_a \mathbb D - \mathfrak{f}_a{}^b K_b \ , 
\qquad e_a := e_a{}^m \partial_m ~,
\ee
where
\be \omega_a{}^{bc} := e_a{}^m \omega_m{}^{bc} \ , \quad b_a := e_a{}^m b_m \ , \quad \mathfrak{f}_a{}^b := e_a{}^m \mathfrak{f}_m{}^b \ .
\ee
The covariant derivatives satisfy the same algebra as $P_a$, except for the introduction of curvatures and torsion
\be
[\nabla_a , \nabla_b] = -T_{ab}{}^c \nabla_c - \hf \RM_{ab}{}^{cd} M_{cd}
	- \RD_{ab} \mathbb D - \RK_{ab}{}^c K_c \ ,
\ee
where the curvatures and torsion are given by the form expressions:
\bsubeq
\begin{align}
T^a&= \rd e^a+ e^a \wedge b + e^b \wedge \omega_b{}^a \ , \\
\RM^{ab} &= \rd \omega^{ab} + \omega^{ac} \wedge \omega_c{}^b - 4 e^{[a} \wedge \mathfrak{f}^{b]} \ , \\
\RD &= \rd b + 2 e^a \wedge \mathfrak{f}_a \ , \\
\RK^a &= \rd \mathfrak{f}^a - \mathfrak{f}^a \wedge b + \mathfrak{f}^b \wedge \omega_b{}^a \ .
\end{align}
\esubeq
In order to define the spin connection (as a composite object) it is necessary to impose some covariant constraint. The appropriate 
constraint is
\be T_{ab}{}^c = 0 \ . \label{DConst1}
\ee
It is clear that the constraint is Lorentz and dilatation invariant, while the conformal invariance may be checked by making use of the 
Jacobi identity
\begin{align}
[ [ K_a , \nabla_b ] , \nabla_c ] &-  [ K_a , \nabla_c ] , \nabla_b ] + [ [ \nabla_b , \nabla_c ] , K_a ] = 0
	\non\\
	&\implies [K_a , [\nabla_b , \nabla_c ] ] = 2 [ [ K_a , \nabla_{[b} ] , \nabla_{c]} ]~.
\end{align}
The right hand side is identically zero as a result of the conformal algebra, so that
\be [K_a , [\nabla_b , \nabla_c ] ] = 0 \ . \label{N0ConfPrim}
\ee
${}$From here we see that $K_a T_{bc}{}^d = 0$.\footnote{Torsion has dimension $1$ and $K_a$ carries dimension $-1$; therefore, $K_a T_{bc}{}^d$ has nothing 
of lower dimension to transform into.} As a result the constraint \eqref{DConst1} is
conformally covariant.

The $K$-gauge transformation of $b_a$ is
\be
\d_K(\L)\, b_a = -2 \L_a \ .
\ee
It is clear that the $K$-gauge transformations can be completely used up to make the gauge choice
\be b_a = 0 \ .
\ee
In what follows we make use of this gauge choice.

It is necessary to constrain the curvatures to correspond to the structure of conformal gravity. Now
constraining the torsion
\be
T_{ab}{}^c = -\cC_{ab}{}^c + 2 \omega_{[ab]}{}^c 
\ , \quad \cC_{ab}{}^{c} = - 2 e_a{}^m e_b^n \partial_{[m} e_{n]}{}^c = 2 e_{[a} e_{b]}{}^m e_m{}^c
\ee
to vanish \eqref{DConst1} allows one to solve for the Lorentz connection in the usual way,
\be 
\omega_{abc} = \frac{1}{2} (\cC_{abc} - \cC_{acb} - \cC_{bca}) 
\ .
\ee

Next we note that the Lorentz curvature is given by
\be \RM_{ab}{}^{cd} = \cR_{ab}{}^{cd} + 8 \d_{[a}^{[c} \mathfrak{f}_{b]}{}^{d]} \ ,
\ee
where
\be
\cR_{ab}{}^{cd} := 2 e_{[a}{}^m e_{b]}{}^n \partial_m \omega_{n}{}^{cd}
	- 2 \omega_{[a}{}^{cf} \omega_{b]f}{}^{d}
\ee
is the standard Riemann tensor constructed from $\omega$. Since we want the special conformal connection to be a composite field we impose the 
conformal gravity constraint\footnote{From the transformations of the curvatures in the last section it is easy to see that when the torsion vanishes
it holds that $K_e \RM_{abcd} = 0$, which makes eq. \eqref{confConstN=0} a conformally invariant
constraint.}
\be \eta^{bd} \RM_{abcd} = 0 \ . \label{confConstN=0}
\ee
This constraint gives
\be \mathfrak{f}_{ab} = - \frac{1}{2 (D - 2)} \cR_{ab} + \frac{1}{4 (D - 1) (D - 2)} \eta_{ab} \cR \ ,
\ee
where
\be \cR_{ac} := \eta^{bd} \cR_{abcd} \ , \quad \cR := \eta^{ab} \cR_{ab} \ .
\ee
Putting our solution for $\mathfrak{f}_{ab}$ into our expression for $\RM_{ab}{}^{cd}$ leads us to the result that $\RM_{ab}{}^{cd}$ coincides with the conformal 
Weyl tensor
\be \RM_{abcd} = C_{abcd} = \cR_{abcd} - \frac{2}{D-2} (\eta_{a[c} \cR_{d] b} - \eta_{b [c} \cR_{d] a}) + \frac{2}{(D - 1)(D - 2)} \cR \eta_{a [c} \eta_{d] b} \ .
\ee
Furthermore, since $\mathfrak{f}_{ab}$ is symmetric we also have
\be \RD_{ab} = 4 \mathfrak{f}_{[ab]} = 0 \ .
\ee

We can also infer information about the conformal curvature $\RK_{ab}{}^c$.
Due to the constraint \eqref{DConst1} the Bianchi identity
\be \nabla_{[a} \nabla_b \nabla_{c ]} = 0
\ee
may be expanded as
\begin{align}
0 =& \ \hf \nabla_{[a} \RM_{bc]}{}^{de} M_{de} + 2 \RK_{[ab}{}^d M_{c] d}
	- \RM_{[abc]}{}^d \nabla_d - \RD_{[ab} \nabla_{c]} \non\\
	&+ \nabla_{[a} \RD_{bc]} \mathbb D - 2 \RK_{[abc]} \mathbb D + \nabla_{[a} \RK_{bc]}{}^d K_d \non\\
=& \  \hf \nabla_{[a} \RM_{bc]}{}^{de} M_{de} + 2 \RK_{[ab}{}^d M_{c] d} - \RM_{[abc]}{}^d \nabla_d
	\non \\ & \quad
	- 2 \RK_{[abc]} \mathbb D + \nabla_{[a} \RK_{bc]}{}^d K_d \ ,
\end{align}
where we used the fact that $\RD = 0$. This result leads to the identities
\begin{subequations}
\begin{align}
\RK_{[abc]} &= 0 \ , \label{CGDBI1} \\ 
\RM_{[abc]}{}^d &= 0 \ , \label{CGDBI2} \\ 
\nabla_{[a} \RK_{bc]}{}^d &= 0 \ , \label{CGDBI3} \\
\nabla_{[a} \RM_{bc ]}{}^{de} - 4 \RK_{[ab}{}^{[d} \d^{e]}_{c]} &= 0 \label{CGDBI4} \ . 
\end{align}
\end{subequations}
Contracting $c$ with $d$ in eq. \eqref{CGDBI4}, and using the constraint \eqref{confConstN=0}, gives
\be \hf \nabla_c \RM_{ab}{}^{c e} + (D - 3) \RK_{ab}{}^e - 2 \RK_{c [a}{}^c \d^e_{b]} = 0 \ .
\ee
From here we deduce that for $D \geq 3$ we have 
\be \RK_{a b}{}^b = 0 \label{traceN0W}
\ee
and
\be 2 (D-3) \RK_{abc} = \nabla^d \RM_{abcd} = \nabla^d C_{abcd} \ .
\ee
Thus all the curvatures may be expressed in terms of the Weyl tensor $C_{abcd}$ for $D \geq 4$. Therefore, the 
vanishing of the Weyl tensor $C_{abcd}$ implies the vanishing of all the conformal gravity curvatures and hence 
conformal flatness.

The $D = 3$ case is special because the Weyl tensor (the traceless part of $\RM_{abcd}$)
vanishes for the choice of $\omega$ which solves 
the torsion constraint \eqref{DConst1}.\footnote{$3D$ 
has the unique property that the Riemann tensor is completely determined by the 
Ricci tensor.} Due to the constraint \eqref{confConstN=0} it must also correspond to
the traceless part of $\RM_{abcd}$. Thus we automatically have $\RM_{abcd} = 0$.

In $3D$ the commutator of two covariant derivatives only involves the special conformal connection\footnote{$\RK_{ab}{}^c$ is a conformal primary, $K_a \RK_{bc}{}^d = 0$, 
as a result of eq. \eqref{N0ConfPrim}.}
\be
[\nabla_a , \nabla_b] = -\RK_{ab}{}^c K_c \ .
\ee
One can show that
\begin{align}
\RK_{mn}{}^c &:= e_m{}^a e_n{}^b \RK_{ab}{}^c = 2 \partial_{[m} \mathfrak{f}_{n]}{}^c
	+ \omega_{[m}{}^{bc} \mathfrak{f}_{n] b} \non\\
	&= 2 \cD_{[m} \mathfrak{f}_{n]}{}^c \ ,
\end{align}
where we introduce
the Lorentz-covariant derivative
\be 
\cD_m = \partial_m - \hf \omega_m{}^{ab} M_{ab} \ , \quad \cD_a := e_a{}^m \cD_m \ .
\ee
Since the torsion vanishes, the curvature may also be written as
\begin{align}
- \hf W_{abc} := \RK_{abc}
= 2 \cD_{[a} \mathfrak{f}_{b] c} = -\cD_{[a} \cR_{b] c} - \frac{1}{4} \eta_{c [a} \cD_{b]} \cR \ ,
\end{align}
where $W_{abc}$ is the Cotton tensor. Furthermore, it is easy to see that when the Cotton tensor vanishes the space is conformally flat.

Due to the symmetry properties satisfied by the Cotton tensor,\footnote{Keep in mind the eqs. \eqref{CGDBI1} and \eqref{traceN0W}.}
\be W_{abc} = - W_{bac} \ , \quad W_{[abc]} = 0 \ , \quad W_{ab}{}^b = 0 \ ,
\ee
we can instead view it as a traceless symmetric rank 2 tensor
\be
 W_{ab} := \hf \eps_a{}^{cd} W_{cdb} \ , \quad  W_{ab} =  W_{ba} \ , \quad  W^a{}_a = 0 \ .
\ee
The Cotton tensor also satisfies a divergenceless condition as a result of eq. \eqref{CGDBI3}
\be
\nabla^b  W_{ab} = 0 \ .
\ee

Note that we could have also chosen $b_m \neq 0$. However, in this case
$\RD_{ab}$ would still vanish because it is invariant under the $K$-gauge 
transformations, $K_c \RD_{ab} = 0$. In fact, in order to derive the geometry
all that is required is to impose the constraints
\be T_{ab}{}^c = 0 \ , \quad \RM_{abcd} = 0 \ . \label{constraintsN=0}
\ee
Considering the term appearing in front of the covariant derivative in the Bianchi identity
\be \nabla_{[a} \nabla_b \nabla_{c]} = 0 \ ,
\ee
we see that under the constraints \eqref{constraintsN=0} $\RD_{ab}$ vanishes also.
The Cotton tensor is again given by the only surviving 
curvature, the special conformal curvature.
We note that these constraints are conformally invariant, and so the composite
expressions for $\omega_{abc}$ and $f_{ab}$, which depend on $b_m$ in general, retain
their original transformation laws.


\section{$\cN$-extended conformal supergravity}\label{confSUGRA} 

We saw in the last subsection that in the
conformal gravity approach the covariant derivative algebra may 
be expressed in terms of a single primary
superfield: the Weyl tensor for
$D \geq 4$ and the Cotton tensor in
$D=3$.
Therefore in the $3D$ $\cN$-extended case we look for a formulation in which the entire covariant derivative algebra is  
expressed in terms of a single primary superfield, the $\cN$-extended super Cotton tensor. A feature of such a setting is that 
the vanishing of the super Cotton tensor implies trivially that the space is conformally flat.

The super Cotton tensor possesses a different index structure for various values of $\cN$. In the bosonic case,  $\cN = 0$, the  Cotton tensor may be 
expressed in terms of spinor indices as
\be 
W_{\a\b\g\d} := (\g^a)_{\a\b} (\g^b)_{\g\d} W_{ab} =W_{(\a\b\g\d)} \ , 
\ee
which is totally symmetric since $W_{ab}$ is both symmetric and traceless. 
The super Cotton tensors for $\cN = 1$ 
and $\cN = 2$ are described by superfields $W_{\a\b\g} = W_{(\a\b\g)}$ and $W_{\a\b} = W_{(\a\b)}$ and were given in  
\cite{KT-M12}
and \cite{Kuzenko12} respectively. 
For $\cN > 3$ it is known \cite{HIPT} that the super Cotton tensor may be described by 
a totally antisymmetric $\rm SO(\cN)$ superfield $W^{IJKL} = W^{[IJKL]}$, 
while for $\cN = 3$ we will see that the super Cotton 
tensor is described by a real spinor superfield $W_\a$.

For the known formulations of conformal superspace in $4D$ the constrained geometry describing conformal supergravity 
surprisingly takes a simple form \cite{ButterN=1, ButterN=2}, despite gauging the entire structure group.
More precisely, the  curvature structure of the theory resembles super Yang-Mills.
As we will demonstrate below, the corresponding ansatz in $3D$ 
turns out to be a very economical means of constraining the curvatures of the theory.
In what follows we will proceed case by case with increasing values of $\cN$.


\subsection{The $\cN = 1$ case}

We begin by first considering the $\cN = 1$ case. It is necessary to constrain the curvatures so as to describe conformal supergravity. We constrain 
the curvatures by
\be \{ \nabla_\a , \nabla_\b \} = 2 \ri \nabla_{\a\b} \ .
\ee
It then follows from the Bianchi identities that the
remaining commutation relations may be written entirely in terms of the operator
\begin{align} W_\a =& \ W(P)_\a{}^a \nabla_a + W(Q)_\a{}^\b \nabla_\b + \hf W(M)_\a{}{}^{ab} M_{ab} \non\\
&+ W(\mathbb D)_\a \mathbb D + W(K)_\a{}^a K_a + W(S)_\a{}^\b S_\b  \ . \label{OpN1}
\end{align}
The remaining commutation relations are
\begin{align}
[ \nabla_a , \nabla_\a ] &= - \hf (\g_a)_\a{}^\b W_\b \ , \\
[\nabla_a , \nabla_b] &= \frac{\ri}{4} \eps_{abc} (\g^c)^{\a\b} \{ \nabla_\a , W_\b \} \ ,
\end{align}
where $W_\a$ must satisfy the Bianchi identity
\be \{ \nabla^\a , W_\a \} = 0 \ . \label{WBI}
\ee
Moreover, as a result of the Jacobi identities, $W_\a$ must be of dimension-3/2 and a conformal primary:
\be [\mathbb D , W_\a] = \frac{3}{2} W_\a \ , \quad \{ S_\a , W_\b \} = 0 \ . \label{Wconditions}
\ee

We now make the following simple ansatz for $W_\a$\footnote{We are motivated by the fact that the torsion and Lorentz and dilatation 
curvatures vanish in the bosonic case.}
\be
W_\a = W(K)_\a{}^a K_a + W(S)_\a{}^\b S_\b \ .
\ee
Then the Bianchi identity \eqref{WBI} implies
\be W(S)_\a{}^\b = 0 \ , \quad
W_{\a\b\g} := W(K)_{\a }{}^{a}(\g_a)_{\b\g} = W_{(\a\b\g)} \label{zeroCN1}
\ee
and the conformally invariant constraint
\be
\nabla^\a W_{\a \b\g} = 0 \ . \label{divLessN1}
\ee
The conditions \eqref{Wconditions} give
\be \mathbb D W_{\a\b\g} = \frac{5}{2} W_{\a\b\g} \ , \quad S_\d W_{\a\b\g} = 0 \ .
\ee

The covariant derivative algebra takes the simple form
\begin{subequations} \label{N=1Algebra}
\begin{align} \{ \nabla_\a , \nabla_\b \} &= 2 \ri \nabla_{\a\b} \ , \\
[ \nabla_a , \nabla_\a ] &= \frac{1}{4} (\g_a)_\a{}^\b W_{\b \g\d} K^{\g\d} \ , \\
[\nabla_a , \nabla_b] &= - \frac{\ri}{8} \eps_{abc} (\g^c)^{\a\b} \nabla_\a W_{\b\g\d} K^{\g\d} - \frac{1}{4} \eps_{abc} (\g^c)^{\a\b} W_{\a\b\g} S^\g \label{N=1Algebra.3}\ .
\end{align}
\end{subequations}
The above algebra has the property that it may be written in terms of a
primary superfield $W_{\a\b\g}$ with the symmetry properties 
of the $\cN = 1$ super Cotton tensor.
In particular, we see that the only curvatures in \eqref{N=1Algebra.3} which arise
in the algebra are $R(K)_{ab}{}^c$ and $R(S)_{ab}{}^\gamma$, which
should correspond to the component Cotton and Cottino tensors.
In section \ref{Degauging} we will see that $W_{\a\b\g}$ is indeed
proportional to the super Cotton tensor in the formulation of \cite{KLT-M11}.


\subsection{The $\cN = 2 $ case}

In the $\cN = 2$ case we take\footnote{The antisymmetric tensors $\eps^{IJ} = \eps_{IJ}$ are normalized as $\eps^{ 1 2} = \eps_{1  2} = 1$.}
\be \{ \nabla_\a^I , \nabla_\b^J \} = 2 \ri \d^{IJ} \nabla_{\a\b} + 2 \ri \eps_{\a\b} \eps^{IJ} W \ ,
\ee
where
\begin{align} W =& \ W(P)^a \nabla_a + W(Q)^\a_I \nabla_\a^I + \hf W(M)^{ab} M_{ab} + W(\mathbb D) \mathbb D \non\\
&+ W(K)^a K_a + W(S)^\a_I S_\a^I + \hf W(N)^{IJ} N_{IJ} \ .
\end{align}
The remaining commutation relations are
\begin{align} [\nabla_a , \nabla_\b^J ] &= - \eps^{JK} (\g_a)_\b{}^\g [ \nabla_{\g K} , W ] \ , \\
[\nabla_a , \nabla_b] &=  \frac{\ri}{4} \eps_{abc} (\g^c)^{\g\d} \eps^{KL} \{ \nabla_{\g K} , [\nabla_{\d L} , W] \} \ ,
\end{align}
with $W$ satisfying the Bianchi identity
\be \eps^{K (I} \{ \nabla^{\g J)}, [ \nabla_{\g K}, W ] \} = 0 \ . \label{N2BI}
\ee
Moreover, $W$ must be of dimension-$1$ and a conformal primary:
\be [\mathbb D , W] = W \ , \quad [ S^I_\a , W ] = 0 \ . \label{N2Wcon}
\ee 

The Bianchi identity \eqref{N2BI} may be solved by the ansatz
\be W = W(K)^a K_a \ .
\ee
Introducing the notation $W_{\a\b} := W(K)^a(\g_a)_{\a\b}$,
we find the following conformally invariant constraint
\be
\nabla^{\a I} W_{\a\b} = 0 \ , \quad 
\label{divLessN2}
\ee
while the conditions \eqref{N2Wcon} give
\be \mathbb D W_{\a\b} = 2 W_{\a\b} \ , \quad S^K_\g W_{\a\b} = 0 \ .
\ee
Hence $W_{\a\b}$ is a primary superfield of dimension-$2$. We will verify in section \ref{Degauging} that $W_{\a\b}$ 
corresponds to the $\cN = 2$ super Cotton tensor.

We then find the algebra to be
\begin{subequations} \label{N=2Algebra}
\begin{align} \{ \nabla_\a^I , \nabla_\b^J \} &= 2 \ri \d^{IJ} \nabla_{\a\b} - \ri \eps^{IJ} \eps_{\a\b} W_{\g\d} K^{\g\d} \ , \\
[ \nabla_a , \nabla_\b^J ] &= \hf (\g_a)_\b{}^\g \eps^{JK} \nabla_{\g K} W^{\a\d} K_{\a\d} + \ri (\g_a)_{\b\g} \eps^{JK} W^{\g\d} S_{\d K} \ , \\
[\nabla_a , \nabla_b] &= - \frac{\ri}{8} \eps_{abc} (\g^c)^{\g\d} \Big( \eps^{KL} (\nabla_{\g K} \nabla_{\d L} W_{\a\b} K^{\a\b} + 4 \ri \nabla_{\g K} W_{\d\b} S^\b_L) - 8 W_{\g\d} \cJ \Big) \ , 
\end{align}
\end{subequations}
where we conveniently introduce the U(1) generator $\cJ$ defined by
\be N_{KL} = \ri \eps_{KL} \cJ \ , \quad \cJ := - \frac{\ri}{2} \eps^{KL} N_{KL} \ .
\ee
The operator $\cJ$ acts on the covariant derivatives as
\be [\cJ , \nabla_\a^I] = - \ri \eps^{IJ} \nabla_{\a J} \ .
\ee

It is often easier to work in a complex basis for the spinor covariant derivatives:
\be \nabla_\a = \frac{1}{\sqrt{2}} (\nabla_\a^{1} - \ri \nabla_\a^{2}) \ , \quad \bar{\nabla}_\a = - \frac{1}{\sqrt{2}} (\nabla_\a^{1} + \ri \nabla_\a^{2}) \ ,
\ee
with definite U(1) charges:
\be [\cJ , \nabla_\a] = \nabla_\a \ , \quad [\cJ , \bar{\nabla}_\a] = - \bar{\nabla}_\a \ .
\ee
The $\rm SO(2)$ connection and curvature then take the form
\be \hf \Phi_A{}^{KL} N_{KL} = \ri \Phi_A \cJ \ , \quad \hf \RN_{AB}{}^{KL} N_{KL} = \ri \RJ_{AB} \cJ \ .
\ee
The conjugation rule in the complex basis is
\be (\nabla_\a F)^* = (-1)^{\eps(F)} \bar{\nabla}_\a \bar{F} \ ,
\ee
where $F$ is a complex superfield and $\bar{F} = (F)^*$ is its complex conjugate.

In the new basis $(\nabla_\a , \bar{\nabla}_\a)$, the covariant derivative algebra \eqref{N=2Algebra} takes the form
\begin{subequations} \label{N=2AlgebraCB}
\begin{align} 
\{ \nabla_\a , \nabla_\b \} &= 0 \ , \quad \{ \bar{\nabla}_\a , \bar{\nabla}_\b \} = 0 \ , \\
\{ \nabla_\a , \bar{\nabla}_\b \} &= - 2 \ri \nabla_{\a\b} - \eps_{\a\b} W_{\g\d} K^{\g\d}\ , \\
[\nabla_a , \nabla_\b] &= \frac{\ri}{2} (\g_a)_\b{}^\g \nabla_\g W^{\a\d} K_{\a\d} - (\g_a)_{\b\g} W^{\g\d} \bar{S}_\d \ , \\
[\nabla_a , \nabla_b] &= - \frac{\ri}{8} \eps_{abc} (\g^c)^{\g\d} \Big( \ri [\nabla_\g , \bar{\nabla}_\d] W_{\a\b} K^{\a\b} + 4 \bar{\nabla}_\g W_{\d\b} \bar{S}^\b + 4 \nabla_\g W_{\d \b} S^\b \\
&\qquad - 8 W_{\g\d} \cJ \Big) \ , 
\end{align}
\end{subequations}
where we define
\be S_\a := \frac{1}{\sqrt{2}} (S_\a^{1} + \ri S_\a^{2}) \ , \quad \bar{S}_\a := \frac{1}{\sqrt{2}} (S_\a^{1} - \ri S_\a^{2}) \ .
\ee
In the complex basis the generators act on the covariant derivatives as
{\allowdisplaybreaks\begin{subequations} \label{gensCB}
\begin{gather}
[M_{\a\b} , \nabla_\g] = \eps_{\g(\a} \nabla_{\b)} \ , \quad [M_{\a\b} , \bar{\nabla}_\g] = \eps_{\g(\a} \bar{\nabla}_{\b)} \ , \\
[\mathbb D, \nabla_\a] = \hf \nabla_\a \ , \quad [\mathbb D, \bar{\nabla}_\a] = \hf \bar{\nabla}_\a \ , \\
 [\cJ , \nabla_\a] = \nabla_\a \ , \quad [\cJ , \bar{\nabla}_\a] = - \bar{\nabla}_\a \ , \\
\{ S_\a , S_\b \} = 0 \ , \quad \{ \bar{S}_\a , \bar{S}_\b \} = 0 \ , \quad \{ S_\a , \bar{S}_\b \} = 2 \ri K_{\a\b} \ , \\
[S_\a , K_b] = 0 \ , \\
[M_{\a\b} , S_\g] = \eps_{\g(\a} S_{\b)} \ , \quad [M_{\a\b} , \bar{S}_\g] = \eps_{\g(\a} \bar{S}_{\b)} \ , \\
[\mathbb D, S_\a] = - \hf S_\a \ , \quad [\mathbb D, \bar{S}_\a] = - \hf \bar{S}_\a \ , \\
 [\cJ , \bar{S}_\a] = \bar{S}_\a \ , \quad [\cJ , S_\a] = - S_\a \ , \\
[K_a , \nabla_\a ] = - \ri (\g_a)_\a{}^\b \bar{S}_\b \ , \quad [K_a , \bar{\nabla}_\a ] = \ri (\g_a)_\a{}^\b S_\b \ , \\
[\bar{S}_\a , \nabla_a] = \ri (\g_a)_\a{}^\b Q_{\b} \ , \quad [S_\a , \nabla_a] = - \ri (\g_a)_\a{}^\b \bar{Q}_{\b} \ , \\
\{ \bar{S}_\a , \nabla_\b \} = 0 \ , \quad \{ S_\a , \bar{\nabla}_\b \} = 0 \ , \\
\{ \bar{S}_\a , \bar{\nabla}_\b \} = - 2 \eps_{\a\b} \mathbb D + 2 M_{\a\b} - 2 \eps_{\a\b} \cJ \ , \quad \{ S_\a , \nabla_\b \} = 2 \eps_{\a\b} \mathbb D - 2 M_{\a\b} - 2 \eps_{\a\b} \cJ \ .
\end{gather}
\end{subequations}}
One may compare the equations \eqref{gensCB} with the algebra given in four-dimensional $\cN = 1$ conformal superspace \cite{ButterN=1}.


\subsection{The $\cN = 3$ case}

In the $\cN = 3$ case we take
\be \{ \nabla_\a^I , \nabla_\b^J \} = 2 \ri \d^{IJ} \nabla_{\a\b} + 2 \ri \eps_{\a\b} W^{IJ} \ ,
\ee
where we require $W^{IJ}$ to have dimension $1$ and be a conformal primary
\be [\mathbb D , W^{IJ}] = W^{IJ} \ , \quad [S_\a^I , W^{JK}] = 0 \ .
\ee
We find the remaining commutation relations
\begin{align} [\nabla_a , \nabla_\a^I ] &= - \frac{1}{2} (\g_a)_\a{}^\b [ \nabla_{\b J} , W^{IJ} ] \ , \\
[\nabla_a , \nabla_b] &= \frac{\ri}{12} \eps_{abc} (\g^c)^{\a \b} \{ \nabla_{\a K} , [ \nabla_{\b L} , W^{KL}] \} \ ,
\end{align}
and the Bianchi identity
\be [ \nabla_\g^{I}, W^{JK} ] = [ \nabla_\g^{[I}, W^{JK]} ] - \frac{1}{2} (\d^{IJ} [ \nabla_{\g L}, W^{KL} ] - \d^{IK} [ \nabla_{\g L}, W^{JL} ] ) \ . \label{N3BI}
\ee

Based on our experience with the previous cases we expect that the covariant derivative algebra should be expressed entirely in terms 
of the $\cN = 3$ super Cotton tensor, $W_\a$. 
We therefore conjecture
\be W^{IJ} := \eps^{IJK} W_K ~, \qquad W^K = \ri W^\g S_\g^K + A (\g^c)^{\a\b} (\nabla_\a^K W_\b ) K_c
\ee
and $A$ is some constant to be determined. Requiring $W^I$ to be a conformal primary fixes the coefficient as
\be W^K = \ri W^\g S_\g^K + \hf (\g^c)^{\g\d} (\nabla_\g^K W_\d) K_c \ .
\ee
Furthermore, the Bianchi identity \eqref{N3BI} is identically satisfied if we demand the conformally invariant 
constraint
\be
\nabla^{\g I} W_\g = 0 \ . \label{divLessN3}
\ee

We find the algebra to be
\begin{subequations} \label{N=3Algebra}
\begin{align} 
\{ \nabla_\a^I , \nabla_\b^J \} &= 2 \ri \d^{IJ} \nabla_{\a\b} - 2 \eps_{\a\b} \eps^{IJL} W^\g S_{\g L} + \ri \eps_{\a\b} (\g^c)^{\g\d} \eps^{IJK} (\nabla_{\g K} W_\d) K_c \ , \\
 [\nabla_a , \nabla_\b^J ] &
 = \eps^{JKL} (\g_a)_{\b\g} \Big[
 \ri 
 W^\g N_{KL} 
 + 
 \ri 
 (\nabla^\g_K W^\d) S_{\d L} 
 \non\\
 &\qquad 
 + \frac{1}{4} 
 (\g^c)_{\d\rho} (\nabla^\g_K \nabla^\d_L W^\rho) K_c \Big] \ , \\
[\nabla_a , \nabla_b] &= 
-\hf 
\eps_{abc} (\g^c)_{\a\b} \eps^{IJK}
\Big[ 
(\nabla^\a_I W^\b) N_{JK} + \frac{1}{2} 
(\nabla^\a_I \nabla^\b_J W^\g ) S_{\g K} \non\\
&\qquad -\frac{\ri}{12} 
(\g^d)_{\g\d} (\nabla^\a_I \nabla^\b_J \nabla^\g_K W^\d) K_d \Big] \ .
\end{align}
\end{subequations}


\subsection{The $\cN > 3$ case}

For the $\cN > 3$ case we again take
\be \{ \nabla_\a^I , \nabla_\b^J \} = 2 \ri \d^{IJ} \nabla_{\a\b} + 2 \ri \eps_{\a\b} W^{IJ}
\ee
and require $W^{IJ}$ to be of dimension-$1$ and a conformal primary
\be [\mathbb D , W^{IJ}] = W^{IJ} \ , \quad [S_\a^I , W^{JK}] = 0 \ .
\ee
Then we find the remaining commutation relations to be
\begin{align} [\nabla_a , \nabla_\a^I ] &= - \frac{1}{(\cN - 1)} (\g_a)_\a{}^\b [ \nabla_{\b J} , W^{IJ} ] \ , \\
[\nabla_a , \nabla_b] &= \frac{\ri}{2 \cN (\cN - 1)} \eps_{abc} (\g^c)^{\a \b} \{ \nabla_{\a K} , [ \nabla_{\b L} , W^{KL}] \} \ ,
\end{align}
where $W^{IJ}$ satisfies the Bianchi identity
\be [ \nabla_\g^{I}, W^{JK} ] = [ \nabla_\g^{[I}, W^{JK]} ] - \frac{1}{\cN - 1} (\d^{IJ} [ \nabla_{\g L}, W^{KL} ] - \d^{IK} [ \nabla_{\g L}, W^{JL} ] ) \ . \label{N>4BI}
\ee
The above algebra and constraints are modeled on those describing a vector multiplet, see appendix \ref{VM}.

Now we expect that the covariant derivative algebra should be expressed entirely in terms 
of the $\cN > 3$ super Cotton tensor, $W^{IJKL}$ and at the lowest dimension we expect it will appear in front of the $\rm SO(\cN)$ generator 
(see \cite{KLT-M11} or section \ref{Degauging}). We therefore 
conjecture that $W^{IJ}$ takes the form
\be W^{IJ} = \hf W^{IJKL} N_{KL} + A (\nabla^\a_K W^{IJKL}) S_{\a L} + B \ri (\g^c)^{\a\b} (\nabla_{\a K} \nabla_{\b L} W^{IJKL}) K_c \ ,
\ee
with $A$ and $B$ some constants to be determined. Requiring $W^{IJ}$ to be a conformal primary fixes the 
coefficients as
\begin{align} W^{IJ} &= \hf W^{IJKL} N_{KL} - \frac{1}{2 (\cN - 3)} (\nabla^\a_K W^{IJKL}) S_{\a L} \non\\
&\quad - \frac{\ri}{4 (\cN - 2) (\cN - 3)} (\g^c)^{\a\b} (\nabla_{\a K} \nabla_{\b L} W^{IJKL}) K_c \ ,
\end{align}
while the Bianchi identity \eqref{N>4BI} for $\cN>4$
 is identically satisfied if we demand the conformally invariant constraint
\be \nabla_{\a}^I W^{JKLP} = \nabla_\a^{[I} W^{JKLP]} - \frac{4}{\cN - 3} \nabla_{\a Q} W^{Q [JKL} \d^{P] I} \ . \label{CSBIN>3}
\ee

In the $\cN=4$ case, the equation \eqref{CSBIN>3} is trivially satisfied,  and instead 
a fundamental Bianchi identity occurs at  dimension-2. 
The super Cotton tensor is equivalently described by a scalar primary superfield in this case, 
 $W^{IJKL}:=\ve^{IJKL}W$,  and eq. \eqref{N>4BI}
is solved by 
\bea
\nabla^{\a I}\nabla_{\a}^JW=\frac{1}{4}\d^{IJ}\nabla^{\a}_P\nabla_{\a}^PW~.
\eea

The algebra of covariant derivatives for $\cN>3$ may be found to be 
\begin{subequations} \label{N>3Algebra}
\begin{align} 
\{ \nabla_\a^I , \nabla_\b^J \} =&\ 2 \ri \d^{IJ} \nabla_{\a\b} + \ri \eps_{\a\b} W^{IJKL} N_{KL} - \frac{\ri}{\cN - 3} \eps_{\a\b} (\nabla^\g_K W^{IJKL}) S_{\g L} \non\\
& + \frac{1}{2 (\cN - 2)(\cN - 3)} \eps_{\a\b} (\g^c)^{\g\d} (\nabla_{\g K} \nabla_{\d L} W^{IJKL}) K_c \ , \\
[\nabla_a , \nabla_\b^J ] =&\ \frac{1}{2 (\cN - 3)} (\g_a)_{\b\g} (\nabla^\g_K W^{JPQK}) N_{PQ} \non\\
& - \frac{1}{2 (\cN - 2) (\cN - 3)} (\g_a)_{\b\g} (\nabla^\g_L \nabla^\d_P W^{JKLP}) S_{\d K} \non\\
& - \frac{\ri}{4 (\cN - 1) (\cN - 2)(\cN - 3)} (\g_a)_{\b\g} (\g^c)_{\d\rho} (\nabla^\g_K \nabla^\d_L \nabla^\rho_P W^{JKLP}) K_c \ , \\
[\nabla_a , \nabla_b] 
=&\    \frac{1}{4 (\cN - 2) (\cN - 3)}   \eps_{abc} (\g^c)_{\a\b} 
\Big( \ri
(\nabla^\a_I \nabla^\b_J W^{PQIJ}) N_{PQ} \non\\
& + \frac{\ri}{ (\cN - 1)} (\nabla^\a_I \nabla^\b_J \nabla^\g_K W^{LIJK}) S_{\g L} \non\\
& + \frac{1}{2 \cN (\cN - 1) } (\g^d)_{\g\d} (\nabla^\a_I \nabla^\b_J \nabla^\g_K \nabla^\d_L W^{IJKL}) K_d \Big) \ .
\end{align}
\end{subequations}

It is worth mentioning that although we considered the $\cN > 3$ case separately, 
its covariant derivative algebra contains information about the 
lower $\cN$ cases. To see this let us consider each value of $\cN$ separately.

For the $\cN = 3$ case we may formally rewrite all terms in the algebra \eqref{N>3Algebra} involving spinor derivatives of $W^{IJKL}$ in terms of $W_\a$,
\be \eps^{IJK} W_\a := - \frac{\ri}{2 (\cN - 3)} \nabla_{\a L} W^{IJKL} \ .
\ee
Then by independently switching off the remaining super Cotton tensor
$W^{IJKL}$
we recover the algebra \eqref{N=3Algebra}.

For the $\cN = 2$ case 
we similarly rewrite all terms involving two or more spinor derivatives in the algebra \eqref{N>3Algebra} in terms of $W_{\a\b}$:
\be
\eps^{IJ} W_{\a \b} := \frac{\ri}{2 (\cN - 2) (\cN - 3)} \nabla_{\a K} \nabla_{\b L} W^{IJKL} \ .
\ee
Independently switching off the remaining terms produces the algebra \eqref{N=2Algebra}.

Finally, the $\cN = 1$ case may be recovered similarly. To do so we introduce $W_{\a\b\g}$ as
\be
W_{\a\b\g} := \frac{\ri}{(\cN- 1)(\cN - 2) (\cN - 3)} \nabla_{(\a K} \nabla_{\b L} \nabla_{\g ) P} W^{JKLP}
\ee
and set to zero the lower dimension fields in the algebra \eqref{N>3Algebra}. This precisely recovers the $\cN = 1$ algebra \eqref{N=1Algebra}. 

Thus we may recover all cases from the algebra \eqref{N>3Algebra}. The corresponding Bianchi identities \eqref{divLessN1}, \eqref{divLessN2} and \eqref{divLessN3} can 
be similarly deduced from the consequences of the Bianchi identity \eqref{CSBIN>3}.

To conclude this section,  
we note that the $\cN=8$ case is special since the super Cotton tensor is a reducible tensor.
We can consistently constrain $W^{IJKL}$ to be self-dual or anti-self-dual.
The resulting conformal supergeometry  may be shown to reduce, 
upon degauging  spelled out in the next section,  
to the $\cN=8$ Weyl supermultiplet described in \cite{Howe:2004ib}.


\section{Degauging $\cN$-extended conformal superspace} \label{Degauging}

Although the conformal superspace constructed in the previous section involves gauging the entire superconformal algebra,\footnote{As in the 
case of superconformal tensor calculus \cite{vanN85,Uematsu,RvanN86,LR89}.} this has traditionally not been the case for conventional superspace formulations. This is 
because it was seen as unnecessary, since with a smaller structure group the local scale and the special conformal transformations may be realized economically
as special gauge transformations, known as super-Weyl transformations. This is exactly the approach adopted in \cite{HIPT,KLT-M11} where 
the $\cN$-extended case in superspace was addressed by gauging $\rm SL(2 , \dsR) \times SO(\cN)$. We will refer 
to the construction of \cite{HIPT,KLT-M11} as $\rm SO(\cN)$ superspace.

In this section we will show how the conventional gauging of \cite{KLT-M11} may be seen to originate within the conformal superspace formulated in the last 
section. We begin with a discussion of some of the salient facts of $\rm SO(\cN)$ superspace and then 
show how the superspace may be derived via gauge-fixing some of the symmetries of conformal superspace. Furthermore, by using the 
degauging procedure of this section we verify our claim that the primary superfields appearing in each of the covariant derivative algebras are 
the corresponding super Cotton tensors. Finally, we derive the super-Weyl transformations of $\rm SO(\cN)$ superspace entirely from our 
conformal superspace.


\subsection{$\rm SO(\cN)$ superspace}

The superspace geometry of \cite{HIPT,KLT-M11} has 
the
structure group $\rm SL(2 , \dsR) \times SO(\cN)$. The covariant derivatives
are given by\footnote{Here, we have slightly altered the conventions of \cite{KLT-M11} by
changing the signs of the connections, torsions and curvatures.}
\be
\cD_A = E_A{}^M \partial_M - \hf \Omega_A{}^{bc} M_{bc} - \hf \Phi_A{}^{PQ} N_{PQ} \ ,
\ee
with the algebra
\be
[\cD_A , \cD_B \} = -T_{AB}{}^C \cD_C - \hf R_{AB}{}^{cd} M_{cd} - \hf R_{AB}{}^{PQ} N_{PQ} \ .
\ee

The torsion is subject to the \emph{conventional} constraints \cite{HIPT}:
\begin{subequations} \label{KLT-MTC}
\begin{align}
T_\a^I{}_\b^J{}^c = -2 \ri \d^{IJ} (\g^c)_{\a\b} \ &, \qquad {\rm (dimension \ 0)} \\
T_\a^I{}_\b^J{}^\g_K = 0 \ , \qquad T_\a^I{}_b{}^c = 0 \ &, \qquad {\rm (dimension \ 1/2)} \\
T_{ab}{}^c = 0 \ , \qquad \eps^{\b\g} T_a{}_\b^{[J}{}_\g^{K]} = 0 \ &. \qquad {\rm (dimension \ 1)}
\end{align}
\end{subequations}
The solution to the constraints \eqref{KLT-MTC} is given in terms of the superfields
\begin{align} W^{IJKL} = W^{[IJKL]} \ , \quad S^{IJ} = S^{(IJ)} \ , \quad C_a{}^{IJ} = C_a{}^{[IJ]} \ , 
\end{align}
which appear at dimension-$1$ in the covariant derivative algebra\footnote{We have placed a prime on the vector covariant derivative since it will 
differ from the one we find from straightforward degauging. 
We have also denoted the super Cotton tensor by $W^{IJKL}$ instead of $X^{IJKL}$.}
\begin{align} \{ \cD_\a^I , \cD_\b^J \} =& \ 2 \ri \d^{IJ} (\g^c)_{\a\b} \cD'_c 
- 2 \ri \eps_{\a\b} C^{\g\d IJ} M_{\g\d} - 4 \ri S^{IJ} M_{\a\b} \non\\ 
&+\Big(\ri \eps_{\a\b} W^{IJKL} - 4 \ri \eps_{\a\b} S^{K[ I} \d^{J] L} + \ri C_{\a\b}{}^{KL} \d^{IJ} - 4 \ri C_{\a\b}{}^{K(I} \d^{J)L} \Big) N_{KL} \ .
\end{align}
The Bianchi identities
imply the dimension-3/2 constraints \cite{KLT-M11}
\begin{subequations} \label{BIarbN}
\begin{align}
\cD_\a^I \cS^{JK} &= 2 \cT_\a{}^{I(JK)} + \cS_\a{}^{(J} \d^{K) I} - \frac{1}{\cN} \cS_\a{}^I \d^{JK} \ , \\
\cD_\a^I C_{\b\g}{}^{JK} &= \frac{2}{3} \ve_{\a(\b} \Big( C_{\g )}{}^{IJK} + 3 \cT_{\g )}{}^{JKI} + 4 (\cD_{\g )}^{[J} \cS) \d^{K] I} + \frac{\cN - 4}{\cN} \cS_{\g )}{}^{[J} \d^{K] I} \Big) \non\\ 
&\quad + C_{\a\b\g}{}^{IJK} - 2 C_{\a\b\g}{}^{[J} \d^{K] I} \ , \\ 
\cD_\a^I W^{JKLP} &= W_\a{}^{IJKLP} - 4 C_\a{}^{[JKL} \d^{P] I} \ ,
\end{align}
\end{subequations}
where the symmetry properties of the superfields $\cT_\a{}^{IJK}$, $C_{\a\b\g}{}^{IJK}$, $C_{\a\b\g}{}^I$ and $W_\a{}^{IJKLP}$ are
\begin{subequations}
\begin{align} \cT_\a{}^{IJK} &= \cT_\a{}^{[IJ] K} \ , \quad \d_{JK} \cT_\a{}^{IJK} = \cT_\a{}^{[IJK]} = 0 \ , \\
C_{\a\b\g}{}^{IJK} &= C_{(\a\b\g)}{}^{IJK} = C_{\a\b\g}{}^{[IJK]} \ , \quad C_{\a\b\g}{}^I = C_{(\a\b\g)}{}^I \ , \\
C_\a{}^{IJK} &= C_\a{}^{[IJK]} \ , \quad W_\a{}^{IJKPQ} = W_\a{}^{[IJKPQ]} \ .
\end{align}
\end{subequations}

The superspace formulation of \cite{HIPT,KLT-M11} describes conformal supergravity since the torsion constraints admit super-Weyl 
transformations. The constraints \eqref{KLT-MTC} can be shown to be invariant under arbitrary super-Weyl transformations of the form \cite{KLT-M11}\footnote{In the case $\cN=8$, 
the super-Weyl transformations were first given in  \cite{Howe:2004ib}.}
\bsubeq \label{superWeyl}
\bea
\d_\s\cD_\a^I&=&
\hf \s\cD_\a^I + (\cD^{\b I}\s) M_{\a\b}+(\cD_{\a J} \s) N^{IJ}
~,
\label{N-sW-1}
\\
\d_\s\cD'_a&=& 
\s\cD'_a 
+ \frac{\ri}{ 2}(\g_a)^{\g\d}(\cD_{\g}^K \s)\cD_{\d K} 
+\ve_{abc}(\cD'^b\s) M^{c} 
\non\\ 
&& 
+ \frac{\ri}{ 16}(\g_a)^{\g\d}([\cD_\g^{K},\cD_\d^{L}]\s) N_{KL}~, 
\label{N-sW-2} 
\eea
where $\s$ is a real unconstrained superfield. This requires the torsion and curvature components to transform as\footnote{Notice that $W^{IJKL}$ is the only 
superfield which transforms homogeneously. For $\cN > 3$ it is the super Cotton tensor.}
\bea 
\d_\s S^{IJ}&=&
\s S^{IJ}
- \frac{\ri}{ 8}[\cD^{\g(I},\cD_{\g}^{J)}]\s
~,
\label{N-sW-S}
\\
\d_\s C_{a}{}^{IJ}&=&
\s C_{a}{}^{IJ}
- \frac{\ri}{ 8}(\g_a)^{\g\d}  [\cD_\g^{[I},\cD_\d^{J]}]\s
~,
\label{N-sW-C}
\\
\d_\s W^{IJKL}&=&\s W^{IJKL}~.
\label{N-sW-W}
\eea
\esubeq

Remarkably, the formulation of \cite{HIPT, KLT-M11} treats all cases
simultaneously
and possesses the simple torsion constraints \eqref{KLT-MTC}.


\subsection{Conventional degauging}

The structure of conformal superspace differs from that of \cite{HIPT, KLT-M11} by the
addition of dilatation and special conformal symmetry in the structure group. To fix 
these additional symmetries we follow the procedure given in \cite{ButterN=1, ButterN=2}.

We first note that under a $K_A$-transformation
the dilatation gauge field $B = E^a B_a + E^\a_I B_\a^I$ transforms as
\begin{align}
\d_{K}(\L) B
	= -2 E{}^a \L_a + 2 E{}^\a_I \L_\a^I \ , \label{GFsuperWeyl}
\end{align}
which permits the gauge choice
\be B_A = 0 \ . \label{GCond}
\ee 
This completely removes the dilatation connection from all the covariant derivatives.

The special conformal connection $\mathfrak{F}^A$ still remains. However, its symmetry
has been fixed and as a result we introduce \emph{degauged}
covariant derivatives with no special conformal connection
\be
\cD_A := \nabla_A + \mathfrak{F}_A{}^B K_B \ ,
\ee
where $\cD_A$ corresponds to the structure group $\rm SL(2 , \dsR) \times SO(\cN)$ and possesses the algebra\footnote{The hatted objects denote the degauged versions of 
the torsion and curvatures.}
\be
[\cD_A , \cD_B \} = -\hat{T}_{AB}{}^C \cD_C - \hf \hat{R}_{AB}{}^{cd} M_{cd}
	- \hf \hat{R}_{AB}{}^{IJ} N_{IJ} 
	~.
\ee
In fact, it is possible to show that up to a redefinition of the degauged 
vector covariant derivatives,
the torsion and curvature correspond to those of \cite{KLT-M11}. To see this we first note that the torsion tensors are related by
\be T^a = \hat{T}^a \ , \quad T^\a_I = \hat{T}^\a_I + \ri E^a \wedge \mathfrak{F}^\b_I (\g_a)_\b{}^\a \ .
\ee
We then see that the torsion is constrained as in eqs. \eqref{KLT-MTC} except that\footnote{This torsion component only vanishes for $\cN = 1$.}
\be \eps^{\b\g} \hat{T}_a{}_\b^{[J}{}_\g^{K]} \neq 0 \ .
\ee
This is due to the fact that the degauged covariant derivatives are defined slightly differently to those of \cite{KLT-M11}. We now turn to explaining this 
point by explicitly deriving the constraints obeyed by the special conformal connection coefficients $\mathfrak{F}_A{}^B$.


\subsection{The degauged special conformal connection}

In the gauge \eqref{GCond} the dilatation curvature is\footnote{We have lowered the index on the $K$-connection as 
$\mathfrak{F}_{A b} = \eta_{bc} \mathfrak{F}_A{}^c$ and $\mathfrak{F}_{A}{}_\b^I = \eps_{\b\g} \d^{IJ} \mathfrak{F}_A{}^\g_J$.}
\be \RD_{AB} = 2 \mathfrak{F}_{AB} (-1)^{\eps_B} - 2 \mathfrak{F}_{BA} (-1)^{\eps_A + \eps_A \eps_B} \ .
\ee
The vanishing of the dilatation curvature constrains the special conformal connection as
\be \mathfrak{F}_{AB} = \mathfrak{F}_{BA} (-1)^{\eps_A \eps_B + \eps_A + \eps_B}
\ee
which implies\footnote{The reason for the chosen coefficients will be clear later.}
\begin{subequations}
\begin{align}
\mathfrak{F}_\a^I{}_\b^J &= - \mathfrak{F}_\b^J{}_\a^I = \ri C_{\a\b}{}^{IJ} - \ri \eps_{\a\b} S^{IJ} \ , \\
\mathfrak{F}_{\a\b,}{}_\g^K &= - \mathfrak{F}_\g^K{}_{,\a\b} = C_{\a\b\g}{}^K + \frac{2}{3} \eps_{\g (\a} C_{\b )}{}^K \ , \\
\mathfrak{F}_{ab} &= \mathfrak{F}_{ba} \ ,
\end{align}
\end{subequations}
where the superfields $S^{IJ}$, $C_{\a\b}{}^{IJ}$, $C_{\a\b\g}{}^I$ and $C_\a{}^I$ satisfy the symmetry properties
\be S^{IJ} = S^{(IJ)} \ , \quad C_{\a\b}{}^{IJ} = C_{(\a\b)}{}^{[IJ]} \ , \quad C_{\a\b\g}{}^I = C_{(\a\b\g)}{}^I \ .
\ee
${}$From here it is possible to derive the degauged covariant derivative algebra 
by computing $[ \cD_A , \cD_B \}$. An efficient way to do this is to consider 
a conformal primary tensor superfield $\Phi$ transforming in some  representation 
of the remainder of the superconformal algebra (compare with \cite{ButterN=2}). For example, 
to determine the dimension-$1$ covariant derivative algebra we consider
\be
\{ \cD_\a^I , \cD_\b^J \} \Phi = \{ \nabla_\a^I , \nabla_\b^J \} \Phi
	+ \mathfrak{F}_\a^I{}^C [K_C , \nabla_\b^J \} \Phi
	+ \mathfrak{F}_\b^J{}^C [K_C , \nabla_\a^I \} \Phi \ .
\ee
Making use of the form of $\mathfrak{F}$ and of the superconformal algebra we find
\begin{align}
\{ \cD_\a^I , \cD_\b^J \} =& \ 2 \ri \d^{IJ} (\g^c)_{\a\b} \cD_c - 2 \ri \eps_{\a\b} C^{\g\d IJ} M_{\g\d} - 4 \ri S^{IJ} M_{\a\b} \non\\ 
&+\Big(\ri \eps_{\a\b} W^{IJKL} - 4 \ri \eps_{\a\b} S^{K[ I} \d^{J] L} - 4 \ri C_{\a\b}{}^{K(I} \d^{J)L} \Big) N_{KL} \ , \non\\
=& \ 2 \ri \d^{IJ} (\g^c)_{\a\b} \cD'_c - 2 \ri \eps_{\a\b} C^{\g\d IJ} M_{\g\d} - 4 \ri S^{IJ} M_{\a\b} \non\\ 
&+\Big(\ri \eps_{\a\b} W^{IJKL} - 4 \ri \eps_{\a\b} S^{K[ I} \d^{J] L} + \ri C_{\a\b}{}^{KL} \d^{IJ} - 4 \ri C_{\a\b}{}^{K(I} \d^{J)L} \Big) N_{KL} \ , \non\\
\end{align}
where
\be \cD'_a = \cD_a - \hf C_a{}^{IJ} N_{IJ} \ . \label{cD'}
\ee
The degauged covariant derivative algebra agrees with the one given in \cite{KLT-M11}, with the vector covariant derivative defined above. The reason for 
the difference in the vector covariant derivative can be attributed to the appearance of the non-zero torsion component,
\be
\eps^{\b\g} \hat{T}_a{}_\b^{[J}{}_\g^{K]} = -2 C_a{}^{JK} \ ,
\ee
which can be removed if one redefines the vector covariant derivative as in eq. \eqref{cD'}.\footnote{In fact, the definition of the vector covariant derivative in \cite{KLT-M11} 
was chosen due to the simpler looking torsion constraints.} Therefore, the degauged version of conformal superspace is constrained in such a way so as to 
correspond to the formulation of \cite{HIPT, KLT-M11}.

The torsion and curvature in \cite{KLT-M11} are constrained by a set of dimension-$3/2$ Bianchi identities. These must follow directly from the degauging procedure. 
To derive these explicitly, we analyze the constraints imposed on the special conformal
curvatures
\begin{subequations} \label{Kcurvs}
\begin{align}
\RS_{AB}{}^\g_K =& \ 2 \cD_{[A} \mathfrak{F}_{B \}}{}^\g_K
	+ \hat{T}_{AB}{}^D \mathfrak{F}_D{}^\g_K
	+ \ri \d_A{}^\d_K \mathfrak{F}_B{}^c (\g_c)_\d{}^\g (-1)^{\eps_B} \non\\
	&- \ri \d_B{}^\d_K \mathfrak{F}_A{}^c (\g_c)_\d{}^\g (-1)^{\eps_B \eps_A+\eps_A} \ , \label{R(K)1} \\
\RK_{AB}{}^c =& \ 2 \cD_{[A} \mathfrak{F}_{B \}}{}^c + \hat{T}_{AB}{}^D \mathfrak{F}_D{}^c
	+ \ri \mathfrak{F}_A{}^\g_K \mathfrak{F}_B{}^{\d K} (\g^c)_{\g \d} (-1)^{\eps_B} \non\\
	&- \ri \mathfrak{F}_B{}^\g_K \mathfrak{F}_A{}^{\d K} (\g^c)_{\g\d} (-1)^{\eps_B \eps_A+\eps_A} \ , \label{R(K)2}
\end{align}
\end{subequations}
which appear in the covariant derivative algebra of the conformal covariant derivatives $\nabla_A$.
We will consider each case in turn.


\subsubsection{The $\cN = 1$ case}
In the $\cN=1$ case the special conformal connection is given by
\begin{subequations}
\begin{align}
\mathfrak{F}_{\a \b} &= - \mathfrak{F}_{\b \a} = -\ri \eps_{\a\b} S \ , \\
\mathfrak{F}_{\a\b,}{}_\g &= - \mathfrak{F}_\g{}_{,\a\b} = C_{\a\b\g} + \frac{2}{3} \eps_{\g (\a} C_{\b )} \ .
\end{align}
\end{subequations}
From the $\cN=1$ algebra \eqref{N=1Algebra}, we find $\RS_{\a\b}{}^\g = 0$, which together
with \eqref{R(K)1} implies
\be
2 \cD_{(\a} \mathfrak{F}_{\b ) \g} - 2 \ri \mathfrak{F}_{\a\b , \g} + 2 \ri \mathfrak{F}_{\g (\a, \b )} = 0 \
\quad \implies \quad C_\a = \cD_\a S~.
\ee
The constraint $\RS_{a \b}{}^\a = 0$ gives
\be
\cD_a \mathfrak{F}_{\b\a} - \cD_\b \mathfrak{F}_{a \a} + \hat{T}_a{}_\b{}^\g \mathfrak{F}_{\g\a} - \ri \mathfrak{F}_{a , \a \b} = 0 \ .
\ee
Then using the degauged torsion
\be
\hat{T}_{a \b}{}^\g = -(\g_a)_\b{}^\g S
\ee
we deduce the constraint
\be \cD^\a C_{\a\b\g} = - \frac{4 \ri}{3} \cD_{\b\g} S \implies \cD_\a C_{\b\g\d} = \cD_{(\a} C_{\b\g\d )} - \ri \eps_{\a (\b} \cD_{\g\d )} S
\ee
and the final expression for the remaining component of $\mathfrak{F}_{AB}$:
\be
\mathfrak{F}_{ab} = -\frac{\ri}{4} (\g_a)^{\a\b} (\g_b)^{\g\d} \cD_{(\a} C_{\b\g\d )}
	+ \frac{\ri}{6} \eta_{ab} \cD^2 S
	+ \eta_{ab} S^2 \ .
\ee
The above results show that we recover the $\cN = 1$ superspace geometry of \cite{KLT-M11}. Moreover, the results for $\mathfrak{F}_{AB}$ are important because they 
enable us to take a superfield expression in conformal superspace and degauge to the corresponding result in 
the superspace formulation of \cite{KLT-M11}.

With the degauging procedure outlined above we can go one step further. First note that the superfield $W_{\a\b\g}$ has the appropriate index structure and dimension 
to correspond to   the $\cN = 1$ supersymmetric super Cotton tensor. To verify this we can derive an expression for $W_{\a\b\g}$ in the degauged superspace and show that 
it is proportional to the expression given in \cite{KT-M12}.
Using the constraint
\be
\RK_{a \a}{}^b = -\frac{1}{4} (\g_a)_\a{}^\b (\g^b)^{\g\d} W_{\b\g\d}
\ee
and the corresponding definition for $\RK_{a \a}{}^b$ in \eqref{R(K)2}, we find
\be
\frac{1}{4} (\g_a)_\a{}^\b (\g_b)^{\g\d} W_{\b\g\d} = 
	- \cD_a \mathfrak{F}_{\a b} + \cD_\a \mathfrak{F}_{ab} - \hat{T}_{a \a}{}^\b \mathfrak{F}_{\b b} + 2 \ri \mathfrak{F}_\a{}^\g \mathfrak{F}_a{}^\d (\g_b)_{\g\d} \ ,
\ee
which gives
\be W_{\a\b\g} = - \ri \cD^2 C_{\a\b\g} - 2 \cD_{(\a\b} \cD_{\g)} S - 8 S C_{\a\b\g} \ .
\ee
This is indeed proportional to the super Cotton tensor given in \cite{KT-M12}. 
The divergenceless condition \eqref{divLessN1} reduces to
\be \cD^\a W_{\a\b\g} =0 \ .
\ee

Since the degauged special conformal connection is important for comparing the results of conformal superspace with those derived in the formulation 
of \cite{KLT-M11}, we summarize its components below:
\begin{subequations}
\begin{align}
\mathfrak{F}_{\a \b} &= - \mathfrak{F}_{\b \a} = -\ri \eps_{\a\b} S \ , \\
\mathfrak{F}_{\a\b,}{}_\g &= - \mathfrak{F}_\g{}_{,\a\b} =
	C_{\a\b\g} + \frac{2}{3} \eps_{\g (\a} \cD_{\b )} S \ , \\
\mathfrak{F}_{ab} &=
	- \frac{\ri}{4} (\g_a)^{\a\b} (\g_b)^{\g\d} \cD_{(\a} C_{\b\g\d )}
	+ \frac{\ri}{6} \eta_{ab} \cD^2 S
	+ \eta_{ab} S^2 \ .
\end{align}
\end{subequations}


\subsubsection{The $\cN > 1$ case}

 For $\cN >  1$ we will need the lowest dimension component of \eqref{R(K)1},
\be \RS_{\a}^I{}_\b^J{}_\g^K = \cD_\a^I \mathfrak{F}_\b^J{}_\g^K + \cD_\b^J \mathfrak{F}_\a^I{}_\g^K - 2 \ri \d^{IJ} \mathfrak{F}_{\a\b,}{}_\g^K 
+ \ri \d^{I K} \mathfrak{F}_{\a \g}{}_{,}{}^J_\b + \ri \d^{J K} \mathfrak{F}_{\b \g}{}_{,}{}_\a^I \label{arbNR3/2}
\ee
and the constraints $\RS^I_{(\a}{}_{\b )}^J{}_\g^K = \RS_\a^{(I}{}_\b^{J)}{}_\g^K = 0$, which hold for arbitrary $\cN$. First we decompose $\cD_\a^I \cS^{JK}$ and $\cD_\a^I C_{\b\g}{}^{JK}$  as:
\begin{align}
\cD_\a^I \cS^{JK} =& \ \cS_\a{}^{IJK} + 2 \cT_\a{}^{I(JK)} + \cS_\a{}^{(J} \d^{K ) I} - \frac{1}{\cN} \cS_\a{}^I \d^{JK} \ , \non\\
\cD_\a^I C_{\b\g}{}^{JK} =& \  C_{\a\b\g}{}^{IJK} + 2 \cT_{\a\b\g}{}^{I[JK]}- 2 \tilde{C}_{\a\b\g}{}^{[J} \d^{K] I} \non\\
&+ \frac{2}{3} \eps_{\a(\b}\Big( C_{\g )}{}^{IJK} + 2 D_{\g)}{}^{I[JK]} + \cT_{\g )}{}^{[J} \d^{K] I} \Big) \ ,
\end{align}
where we define $\cS^{IJ}$ by the decomposition
\be S^{IJ} = \cS \d^{IJ} + \cS^{IJ} \ , \quad \cS = \frac{1}{\cN} \d_{IJ} S^{IJ} \ , \quad \d_{IJ} \cS^{IJ} = 0
\ee
and we introduce superfields which satisfy the properties
\begin{subequations}
\begin{gather}
\cS_\a{}^{IJK} = \cS_\a{}^{(IJK)} \ , \quad \d_{JK} \cS_\a{}^{IJK} = 0 \ , \\
\cT_\a{}^{IJK} = \cT_\a{}^{[IJ] K} \ , \quad \d_{JK} \cT_\a{}^{IJK} = \cT_\a{}^{[IJK]} = 0 \ , \\
\cD_{\a J} \cS^{IJ} = \frac{(\cN + 2)(\cN -1)}{2 \cN} \cS_\a{}^J \ , \\
C_{\a\b\g}{}^{IJK} = C_{(\a\b\g)}{}^{[IJK]} \ , \quad \tilde{C}_{\a\b\g}{}^I = \tilde{C}_{(\a\b\g)}{}^I \ , \\
C_\a{}^{IJK} = C_\a{}^{[IJK]} \ , \\
\cT_{\a\b\g}{}^{IJK} = \cT_{(\a\b\g)}{}^{(IJ)K} \ , \quad \d_{JK} \cT_{\a\b\g}{}^{IJK} = \cT_{\a\b\g}{}^{(IJK)} = 0 \ , \\
D_\a{}^{IJK} = D_\a{}^{[IJ] K} \ , \quad \d_{JK} D_\a{}^{IJK} = D_\a{}^{[IJK]} = 0 \ , \\
\cD^\g_I C_{\b\g}{}^{IJ} = (\cN - 1) \cT_\b{}^J \ .
\end{gather}
\end{subequations}

Symmetrizing the indices $I$, $J$ and $K$ in eq. \eqref{arbNR3/2} gives
\be 0 = 2 \cD_{(\a}^{(I} \mathfrak{F}_{\b )}^J{}_\g^{K)} - 2 \ri \d^{(IJ} \mathfrak{F}_{\a\b,}{}_\g^{K)} + 2 \ri \d^{(IJ} \mathfrak{F}_{\g (\a ,}{}_{\b)}^{K)} \ ,
\ee
which implies
\be \cD_\b^{(I} S^{JK)} = \d^{(IJ} C_{\b}{}^{K)} \ .
\ee
${}$From here we find
\be \cS_\a{}^{IJK} = 0 \ , \quad C_\b{}^J = \frac{\cN - 1}{\cN} \cS_\b^J + \frac{\cN}{\cN + 2} \cD_\b^J \cS \ .
\ee

Now symmetrizing the indices $\a$, $\b$ and $\g$ in eq. \eqref{arbNR3/2} gives
\be 0 = 2 \cD_{(\a}^{(I} \mathfrak{F}_\b^{J)}{}_{\g )}^K - 2 \ri \d^{IJ} \mathfrak{F}_{(\a\b ,}{}^K_{\g)} + 2 \ri \d^{K (I} \mathfrak{F}_{(\a\b ,}{}_{\g )}^{J)}  \ ,
\ee
and then we deduce
\be \tilde{C}_{\a\b\g}{}^J = C_{\a\b\g}{}^J \ , \quad \cT_{\a\b\g}{}^{IJK} = 0 \ .
\ee

Contracting the indices $\a$ with $\b$ in eq. \eqref{arbNR3/2} 
and using $R(S)_{[a}^I{}_{\b]}^J{}_\g^K=0$ leads to
\be 0 = 2 \cD^{\a [I} \mathfrak{F}_\a^{J]}{}_\g^K + 2 \ri \d^{K [ I} \mathfrak{F}^\a{}_{\g ,}{}_\a^{J]} \ .
\ee
From here it is easy to see that
\be D_\a{}^{IJK} = 0 \ .
\ee
Furthermore, we also find
\be \cD^{\a I} C_{\a\b}{}^{JK} = 3 \cT_{\b}{}^{JKI} + 4 (\cD_\b^{[J} \cS) \d^{K ] I} + \frac{\cN - 4}{\cN} \cS_\b{}^{[J} \d^{K] I} \ .
\ee
Putting all these constraints together precisely recovers the Bianchi identities \eqref{BIarbN} except for the one involving $W^{IJKL}$, which only 
appears for $\cN > 3$. In this case
we can make use of
\be
\RS^{\a I}{}_\a^J{}^\g_K = \frac{2 \ri}{\cN - 3} \nabla^{\g L} W^{IJ}{}_{LK} =  \frac{2 \ri}{\cN - 3} \cD^{\g L} W^{IJ}{}_{LK}
\ee
and eq. \eqref{arbNR3/2} to derive
\be C_\a{}^{IJK} = \frac{1}{\cN - 3} \nabla_{\a L} W^{LIJK} \ .
\ee
Since $W^{IJKL}$ is primary we recover the final Bianchi identity from eq. \eqref{CSBIN>3}
\be \cD_\a^I W^{JKLP} = \cD_\a^{[I} W^{JKLP]} - 4 C_\a{}^{[JKL} \d^{P] I}~.
\ee

So far we have obtained the special conformal connection components:
\begin{subequations}
\begin{align}
\mathfrak{F}_\a^I{}_\b^J &= - \mathfrak{F}_\b^J{}_\a^I = \ri C_{\a\b}{}^{IJ} - \ri \eps_{\a\b} S^{IJ} \ , \\
\mathfrak{F}_{\a\b,}{}_\g^K &= - \mathfrak{F}_\g^K{}_{,\a\b} = C_{\a\b\g}{}^K + \frac{2}{3} \eps_{\g (\a} \Big( \frac{\cN -1}{\cN} \cS_{\b)}{}^J + \frac{\cN}{\cN + 2} \cD_{\b)}^J \cS \Big) \ .
\end{align}
\end{subequations}
The final component $\mathfrak{F}_{ab}$ may also be found by considering each value of $\cN$ separately. We have already shown how to do this 
for $\cN = 1$. Below we illustrate the higher $\cN$ cases and derive the corresponding super Cotton tensors directly from the degauging of 
conformal superspace in the $\cN = 2$ and $\cN = 3$ cases.


\noindent{\bf The $\cN = 2$ case}

In the $\cN = 2$ case the torsion component $C_a{}^{KL}$ takes the form
\be C_a{}^{KL} = C_a \eps^{KL}
\ee
and the remaining constraints become
\begin{subequations}
\begin{align} \cD_\a^I \cS^{JK} &= \cS_\a{}^{(J} \d^{K) I} - \hf \cS_\a{}^I \d^{JK} \ , \\
\cD_\a^I C_{\b\g} &= \eps^{IJ} C_{\a\b\g}{}^J - \frac{1}{3} \eps_{\a ( \b} \eps^{IJ} (4 \cD_{\g )}^J \cS - \cS_{\g)}{}^J ) \ , \\
S^{IJ} &= \cS \d^{IJ} + \cS^{IJ} \ , \quad \cS := \hf \d_{IJ} S^{IJ} \ , \quad \d_{IJ} \cS^{IJ} = 0 \ .
\end{align}
\end{subequations}

To construct both $\mathfrak{F}_{ab}$ and the $\cN = 2$ super Cotton tensor in the formulation of \cite{KLT-M11} we make use of the special conformal curvature
component (see the algebra \eqref{N=2Algebra})
\be
\RK_\a^I{}_\b^J{}^a = \ri \eps^{IJ} \eps_{\a\b} W_{\g\d} (\g^a)^{\g\d} \ .
\ee
Plugging this result into eq. \eqref{R(K)2} for the $\cN = 2$ case
yields
\begin{align}
\ri \eps^{IJ} \eps_{\a\b} W_{\g\d} (\g^a)^{\g\d} =& \ \cD_\a^I \mathfrak{F}_\b^J{}^a + \cD_\b{}^J \mathfrak{F}_\a^I{}^a - 2 \ri \d^{IJ} \mathfrak{F}_{\a\b ,}{}^a \non\\
&- \ri \mathfrak{F}_\a^I{}^\g_K \mathfrak{F}_\b^J{}^{\d K} (\g^a)_{\g\d} - \ri \mathfrak{F}_\b^J{}^\g_K \mathfrak{F}_\a^I{}^{\d K} (\g^a)_{\g\d} \ . \label{N2Cotton}
\end{align}
We then find the super Cotton tensor by antisymmetrizing $\a$ with $\b$ and $I$ with $J$
\begin{align}
W_{\a\b} &= -\frac{\ri}{4} \eps_{IJ} \cD^{\g I} \mathfrak{F}_{\g ,}^J{}_{\a\b} + 2 \cS C_{\a\b} \non\\
&= \frac{\ri}{8} [\cD^\g_I , \cD_\g^I ] C_{\a\b} - \frac{\ri}{4} \eps_{IJ} [\cD_{(\a}^I , \cD_{\b)}^J] \cS + 2 \cS C_{\a\b} \ ,
\end{align}
which reads in the complex basis
\be 
W_{\a\b} = - \frac{\ri}{4} [\cD^\g , {\bar{\cD}}_\g ] C_{\a\b} 
+ \hf [\cD_{(\a} , \bar{\cD}_{\b)}] \cS + 2 \cS C_{\a\b} \ .
\label{5.56}
\ee
This is proportional to the super Cotton tensor constructed in \cite{Kuzenko12}.\footnote{One may 
always choose a super-Weyl gauge $\cS =0$ in which the expression \eqref{5.56} reduces
to that given for the first time by Zupnik and Pak \cite{ZP}.}
 
On the other hand symmetrizing $\a$ with $\b$ and contracting $I$ with $J$ in \eqref{N2Cotton} gives
\begin{align}
\mathfrak{F}_{ab} &= \frac{\ri}{4} \d_{IJ} (\g_a)^{\a\b} \cD_\a^I \mathfrak{F}_\b^J{}_{, b}
	+ \frac{1}{4} \d_{IJ} (\g_a)^{\a\b} (\g_b)_{\g\d} \mathfrak{F}_\a^I{}^\g_K \mathfrak{F}_\b^J{}^{\d K} \non\\
	&= -\frac{\ri}{4} (\g_{(a})^{\a\b} (\g_{b)})^{\g\d} \eps_{IJ} [\cD_\a^I , \cD_\b^J] C_{\g\d}
	- \frac{\ri}{2} \eta_{ab} [\cD^\a_I , \cD_\a^I] \cS \non\\
	& \quad
	- (\g_a)^{\a\b} (\g_b)^{\g\d} C_{\a\g} C_{\b\d}
	+ \eta_{ab} \cS^{IJ} \cS_{IJ}
	+ 2 \eta_{ab} \cS^2 \ .
\end{align}

It should be mentioned that degauging the
constraint \eqref{divLessN2} gives
\be 
\cD^{\a I} W_{\a\b} = 0 \ . 
\ee


\noindent{\bf The $\cN = 3$ case}

In the $\cN = 3$ case the super Cotton tensor appears in the special conformal curvature component
\be
\RS_\a^I{}_\b^J{}^\g_K = 2 \eps_{\a\b} \eps_K{}^{IJ} W^\g \ .
\ee
Therefore, using eq. \eqref{arbNR3/2}, we may derive an expression for the super Cotton tensor in the formulation of \cite{KLT-M11}. We find
\be
2 \eps^{IJK} W_\g = \cD^{\a [I} \mathfrak{F}_\a^{J]}{}^K_\g + \ri \d^{K [I} \mathfrak{F}^\a{}_{\g ,}{}^{J]}_\a \ ,
\ee
which gives
\be
W_\a = \frac{1}{12} \eps_{IJK} \cD^{\b I} \mathfrak{F}_\b^J{}_\a^K = \frac{\ri}{12} \eps_{IJK} \cD^{\b I} C_{\a\b}{}^{JK} \ .
\ee
As a check one can show that $W_\a$ transforms homogeneously under the super-Weyl transformations \eqref{superWeyl}. One can 
also show that the constraint \eqref{divLessN3} degauges to
\be \cD^{\g I} W_\g = 0 \ .
\ee

To construct $\mathfrak{F}_{ab}$ we use the special conformal curvature component (see the algebra \eqref{N=3Algebra})
\be
\RK_\a^I{}_\b^J{}^a = -\ri \eps_{\a\b} (\g^a)^{\g\d} \eps^{IJK} \nabla_{\g K} W_\d \ .
\ee
Making use of eq. \eqref{R(K)2} and the fact that $\RK_{(\a}^{I}{}_{\b) I}{}^a = 0$ gives us
\be
0 = 2 \cD_{(\a}^{I} \mathfrak{F}_{\b) I}{}^a - 6 \ri \mathfrak{F}_{\a\b ,}{}^a
	- 2 \ri \mathfrak{F}_{(\a}^{I}{}^\g_K \mathfrak{F}_{\b ) I}{}^{\d K} (\g^a)_{\g\d}  \ ,
\ee
which yields
\begin{align}
\mathfrak{F}_{ab} =& \
	-\frac{\ri}{12} (\g_{(a})^{\a\b} (\g_{b)})^{\g\d} \cD_{\a I} C_{\b\g\d}{}^I
	- \frac{\ri}{27} \eta_{ab} \cD^\a_I \cS_\a{}^I
	- \frac{\ri}{30} \eta_{ab} \cD^\a_I \cD^I_\a \cS \non\\
	& - \frac{1}{6} (\g_a)^{\a\b} (\g_b)^{\g\d} C_{\a\g}{}^{IJ} C_{\b\d}{}_{IJ}
	+ \frac{1}{3} \eta_{ab} \cS^{IJ} \cS_{IJ} + \eta_{ab} \cS^2 \ .
\end{align}


\noindent{\bf The $\cN > 3$ case}

Similarly, using eq. \eqref{R(K)2} and the fact that
$\RK_{(\a}^{I}{}_{\b) I}{}^a = 0$ for $\cN > 3$ gives us
\be
0 = 2 \cD_{(\a}^{I} \mathfrak{F}_{\b) I}{}^a - 2 \ri \cN \mathfrak{F}_{\a\b ,}{}^a
	+ 2 \ri \mathfrak{F}_{(\a}^{I}{}^\g_K \mathfrak{F}_{\b ) I}{}^{\d K} (\g^a)_{\g\d}  \ ,
\ee
which yields
\begin{align}
\cN \mathfrak{F}_{ab} =& \
	-\frac{\ri}{4} (\g_{(a})^{\a\b} (\g_{b)})^{\g\d} \cD_{\a I} C_{\b\g\d}{}^I
	- \frac{\ri}{6} \frac{\cN - 1}{\cN} \eta_{ab} \cD^\a_I \cS_\a{}^I 
	- \frac{\ri}{6} \frac{\cN}{\cN + 2} \eta_{ab} \cD^\a_I \cD^I_\a \cS \non\\
	&- \hf (\g_a)^{\a\b} (\g_b)^{\g\d} C_{\a\g}{}^{IJ} C_{\b\d}{}_{IJ}
	+ \eta_{ab} \cS^{IJ} \cS_{IJ} + \cN \eta_{ab} \cS^2 \ .
\end{align}

We may summarize the components of $\mathfrak{F}_{AB}$ for $\cN > 1$ as:
\begin{subequations}
\begin{align}
\mathfrak{F}_\a^I{}_\b^J &= - \mathfrak{F}_\b^J{}_\a^I = \ri \eps^{IJ} C_{\a\b} - \ri \eps_{\a\b} S^{IJ} \ , \\
\mathfrak{F}_{\a\b,}{}_\g^K &= - \mathfrak{F}_\g^K{}_{,\a\b} =
	C_{\a\b\g}{}^K + \frac{2}{3} \eps_{\g (\a} \Big( \frac{\cN -1}{\cN} \cS_{\b)}{}^J + \frac{\cN}{\cN + 2} \cD_{\b)}^J \cS \Big) \ , \\
\mathfrak{F}_{ab} =& \ -\frac{\ri}{4 \cN} (\g_{(a})^{\a\b} (\g_{b)})^{\g\d} \cD_{\a I} C_{\b\g\d}{}^I
	- \frac{\ri (\cN - 1)}{6 \cN^2} \eta_{ab} \cD^\a_I \cS_\a{}^I
	- \frac{\ri}{6 (\cN + 2)} \eta_{ab} \cD^\a_I \cD^I_\a \cS \non\\
	&- \frac{1}{2 \cN} (\g_a)^{\a\b} (\g_b)^{\g\d} C_{\a\g}{}^{IJ} C_{\b\d}{}_{IJ}
	+ \frac{1}{\cN} \eta_{ab} \cS^{IJ} \cS_{IJ} + \eta_{ab} \cS^2 \ .
\end{align}
\end{subequations}


\subsection{The conformal origin of the super-Weyl transformations}

In \cite{HIPT,KLT-M11} a formulation for conformal supergravity was given in which
the dilatations and special conformal transformations were not realized manifestly.
As we have just demonstrated, this $\rm SO(\cN)$ superspace can be viewed as a ``degauged''
version of our conformal superspace, where the special conformal symmetry has been
fixed by the gauge condition $B_A = 0$. As we have left the dilatational
symmetry unfixed, it must survive as an additional nonlinear transformation
not residing in the remaining structure group or the general coordinate
transformations. This is precisely the super-Weyl transformation\footnote{This is exactly the same origin as
the Weyl transformation in conformal gravity
as well as the super-Weyl transformations for $4D$ $\cN=1$ \cite{ButterN=1} and
$\cN=2$ \cite{ButterN=2} conformal supergravity.}
and is what ensures 
the superspace formulation of \cite{KLT-M11} describes conformal supergravity.
We may now show explicitly how these super-Weyl transformations originate in the 
degauging of conformal superspace.

Suppose we have gauge fixed the dilatation connection to vanish by using
the special conformal symmetry.
If we now perform a dilatation with parameter $\s$,
we must accompany it with an additional $K_A$ transformation with $\s$-dependent
parameters $\L^A(\s)$ to maintain the gauge $B_A=0$.
With respect to the covariant derivatives this means
\be
	\d_K(\L(\s)) \nabla_A + \d_{\mathbb D} (\s ) \nabla_A
\ee
cannot contain any terms proportional to the dilatation generator $\mathbb D$.
Using eq. \eqref{TransCD}, we find
\be
\L^a(\s) = \hf \cD^a \s \ , \quad \L^\a_I(\s) = - \hf \cD^\a_I \s \ .
\ee
Then the super-Weyl transformations may be simply read off from
\be \d_\s \nabla_A := \d_K(\L(\s)) \nabla_A + \d_{\mathbb D} (\s ) \nabla_A~.
\ee
The super-Weyl transformations of the degauged covariant derivatives $\cD_A$ and 
the special conformal connection can be read from
\be
\d_\s \nabla_A = \d_\s \cD_A - \d_\s \mathfrak{F}_A{}^B K_B~.
\ee
The super-Weyl transformations of $\cD_A$ are found to be
\bsubeq
\bea
\d_\s\cD_\a^I&=&
\hf \s\cD_\a^I + (\cD^{\b I}\s) M_{\a\b}+(\cD_{\a J} \s) N^{IJ}
~,
\\
\d_\s\cD_a&=&
\s\cD_a
+\frac{\ri}{ 2}(\g_a)^{\g\d}(\cD_{\g}^K \s)\cD_{\d K}
+\ve_{abc}(\cD^b\s) M^{c} \ ,
\eea
\esubeq
while the super-Weyl transformation of, for example, $\mathfrak{F}_\a^I{}^\b_J$ is
\be 
\d_\s \mathfrak{F}_\a^I{}^\b_J =
	\s \mathfrak{F}_\a^I{}^\b_J
	- \hf \cD_\a^I \cD^\b_J \s
	+ \frac{\ri}{2} \d^I_J (\g_a)_{\a\b} \cD^a \s
	= \s \mathfrak{F}_\a^I{}^\b_J - \frac{1}{4} [\cD_\a^I , \cD^\b_J ] \s \ . \label{FKsuperWeyl}
\ee

Eq. \eqref{FKsuperWeyl} recovers the super-Weyl transformations of the torsion components in the formulation of \cite{KLT-M11}, while the super-Weyl transformation of the 
degauged vector covariant derivative does not exactly match that of eq. \eqref{superWeyl}, since it does not contain an $\rm SO(\cN)$ contribution. However, the redefined 
vector covariant derivatives
\be \cD'_a = \cD_a - \hf C_a{}^{IJ} N_{IJ}
\ee
possess the appropriate transformation law as expected.

Note that all primary superfields $\Phi$ transform homogeneously
\be
	\d_K(\L(\s)) \Phi + \d_{\mathbb D} (\s ) \Phi = \d_{\mathbb D} (\s ) \Phi = \s \D \Phi \ ,
\ee
where $\D$ is the dimension of $\Phi$
\be \mathbb D \Phi =  \D \Phi \ .
\ee
Therefore we also have
\be \d_\s W^{IJKL} = \s W^{IJKL} \ .
\ee


\section{Discussion and outlook}\label{conclusion} 

In this paper we have constructed a new off-shell formulation 
for  $\cN$-extended conformal supergravity 
in three dimensions, which possesses a number of key properties. 
Firstly, it 
gauges the entire superconformal algebra
without the need to introduce 
the super-Weyl transformations as in the  conventional approach \cite{HIPT,KLT-M11}, 
which is based on the local structure group $\rm SL(2 , \dsR) \times SO(\cN)$.
Secondly, it possesses a simple covariant derivative algebra, 
the structure of which resembles that of the super Yang-Mills algebra.  
Thirdly, the entire algebra of covariant derivatives is expressed in 
terms of a single primary superfield, the $\cN$-extended super Cotton tensor,
which makes our formulation quite geometrical.

Upon
degauging of the local special conformal and $S$-supersymmetry transformations, 
the conformal superspace constructed  in this paper reduces to the conventional 
formulation for conformal supergravity 
\cite{HIPT,KLT-M11}, with the local scale
transformation turning into the super-Weyl 
transformation. This means that there is no need to carry out a thorough component analysis 
to justify that our formalism is indeed suitable to describe conformal supergravity. 

Although the suitability to describe conformal supergravity is justified, it is worth mentioning that 
the 
conformal superspace may be shown to reduce in components to the superconformal 
framework of \cite{vanN85, RvanN86} for the $\cN = 1$ and $\cN = 2$ cases. 
Recall that a supersymmetry transformation with parameter $\xi^\alpha_I$
should be identified with a supergravity gauge transformation \eqref{TransCD}
with $\cK = \xi^\alpha_I \nabla_\alpha^I$.
For the $\cN = 1$ case and with the general ansatz 
\eqref{OpN1}, we may derive the supersymmetry transformations of the connection
fields using \eqref{eq:deltaConn}:\footnote{Per the usual convention,
we have identified $\psi_m{}^\alpha = 2 E_m{}^\alpha|$ and
$\phi_m{}^\alpha = 2 \mathfrak{F}_m{}^\alpha|$.}
\begin{subequations}
\begin{align}
\d_Q e_m{}^a &= - \ri (\xi \g^a \psi_m) - \hf (\xi \g_m)^\b W(P)_\b{}^a| \ , \\
\hf \d_Q \psi_m{}^\a &= (\partial_m - \hf \omega_m{}^{ab} M_{ab} + \hf b_m) \xi^\a
- \hf (\xi \g_m)^\b W(Q)_\b{}^\a| \ , \\
\d_Q b_m &= - (\xi \phi_m) - \hf (\xi \g_m)^\b W(\mathbb D)_\b| \ , \\
\d_Q \omega_m{}^{ab} &= - \eps^{abc} (\g_c)_{\a\b} \xi^\a \phi_m{}^\b 
- \frac{1}{2} (\xi \g_m)^\b W(M)_\b{}^{ab}| \ , \\
\hf \d_Q \phi_m{}^\a &= - \ri (\xi \g_b)^\a f_m{}^b - \hf (\xi \g_m)^\b W(S)_\b{}^\a| \ ,  \\
\d_Q \mathfrak f_m{}^a &= - \hf (\xi \g_m)^\b W(K)_\b{}^a| \ .
\end{align}
\end{subequations}
Now comparing with \cite{vanN85} we see that we must set $W_\a = W(K)_\a{}^a K_a$, 
which was what we used in superspace. Therefore, at least for 
the $\cN = 1$ case the conformal superspace correctly reduces in components 
(up to conventions) to those derived within the superconformal tensor calculus. 
As a result our formulation may provide a useful bridge between the two approaches.

As compared with the conventional formulation \cite{HIPT,KLT-M11}, 
conformal superspace has a larger gauge group. A nontrivial manifestation of this 
enlarged gauge symmetry is a dramatic reduction of dimension-1 curvature tensors.
In the conventional setting, there are several such tensors:  $S^{IJ} = S^{(IJ)}$,
$C_a{}^{IJ} = C_a{}^{[IJ]}$ and $W^{IJKL} = W^{[IJKL]} $.  Their presence makes the 
algebra of covariant derivatives rather involved and somewhat cumbersome 
from the point of view of practical calculations. 
On the other hand, conformal superspace has no dimension-1 curvature  
for the cases $\cN=1,2,3$, while   for $\cN>3$ the entire algebra of covariant derivatives
is constructed entirely in terms of the super Cotton tensor $W^{IJKL}$. 

The fact that the dimension-1 tensors $S^{IJ}$ and $C_a{}^{IJ}$ do not show up in conformal 
superspace is of primary importance for the explicit construction of conformal supergravity actions. 
In section 1, we briefly discussed the method proposed in \cite{KT-M12} 
to construct off-shell supergravity actions in superspace
and the technical difficulty in implementing this method for the case $\cN \geq 2$.
Let us recall that the main technical problem is the existence of a two-parameter freedom to choose 
the vector covariant derivative, eq. \eqref{1.1}. This leads to a two-parameter family of closed 
three-forms  that should be considered as candidates to generate the action for conformal supergravity. 
This two-parameter family has only one true candidate subject to the condition of super-Weyl invariance modulo exact terms. The explicit 
construction of such a form is highly nontrivial. 
In conformal superspace, however, both 
of these problems do not occur by construction.  
Firstly, the vector covariant derivative is uniquely defined. Secondly, the super-Weyl invariance is built in {\it ab initio}. 
The problem of the explicit construction of off-shell actions for conformal supergravities 
will be addressed in an accompanying paper \cite{BKNT-M}.

In this paper we only considered the vector 
supermultiplets in conformal superspace. The rigid $\cN=3$ and $\cN=4$ projective 
hypermultiplets introduced in \cite{KPT-MvU} may naturally be lifted to conformal 
superspace. This will be discussed elsewhere. 

Using the explicit structure of the super Cotton tensors discussed above, 
we can predict the superfield types of 
conformal supergravity prepotentials for  $1\leq \cN \leq 4$
without working out  
unconstrained prepotential formulations for conformal supergravity 
theories.\footnote{On general grounds, 
such formulations should exist at least in the cases $1\leq \cN \leq 4$, 
and they have been constructed in the cases $\cN=1$ \cite{GGRS} and 
$\cN=2$ \cite{Kuzenko12}.}  Indeed, given the off-shell 
action for conformal supergravity,  $ S_{\rm CSG}$, we expect  that 
\bea
W \propto \frac{\d S_{\rm CSG} }{\d H}~,
\eea
with $W$ and $H$ being respectively the super Cotton tensor and the conformal 
prepotential (with all indices suppressed).
This implies that the unconstrained conformal prepotentials should be as follows:
 $H_{\a\b\g} $ 
for $\cN=1$ \cite{GGRS}, $H_{\a\b}$ for $\cN=2$ \cite{ZP,Kuzenko12}, 
$H_\a$ for $\cN=3$, and $H$ for $\cN=4$. Using the harmonic superspace techniques
\cite{GIOS}, one may derive the $\cN=3$ and $\cN=4$ prepotentials by generalizing 
the four-dimensional  $\cN=2$ analysis of \cite{KT} (see also \cite{Siegel-curved}).

In the component approach, there is a remarkable 
(AdS/CFT inspired) construction \cite{NT}
of the $\cN=8$ off-shell conformal supergravity in three dimensions
starting from the $\cN=8$ SO(8) gauge supergravity in four dimensions \cite{deWN}, 
which has an $\rm AdS_4$ solution. It would be interesting to derive a superspace 
analog of this construction. 
\\


\noindent
{\bf Acknowledgements:}\\
The work of DB was supported by
ERC Advanced Grant No. 246974, ``{\it Supersymmetry: a window to
non-perturbative physics}.''
The work of SMK  and JN was supported in part by the Australian Research Council,
project No. DP1096372.  
The work of GT-M and JN was supported by the Australian Research Council's Discovery Early Career 
Award (DECRA), project No. DE120101498.


\appendix


\section{Notation and conventions} \label{NC}

Our conventions for spinors in three spacetime dimensions ($3D$) follow closely those of \cite{KLT-M11}.\footnote{In particular they 
are compatible with the $4D$ two-component spinor formalism used in \cite{WB, Ideas}.} We summarize them here. 

Spinor indices are raised and lowered using the SL(2,${\mathbb R}$) invariant tensors
\bea
\ve_{\a\b}=\left(\begin{array}{cc}0~&-1\\1~&0\end{array}\right)~,\qquad
\ve^{\a\b}=\left(\begin{array}{cc}0~&1\\-1~&0\end{array}\right)~,\qquad
\ve^{\a\g}\ve_{\g\b}=\d^\a_\b
\eea
as follows:
\bea
\psi^{\a}=\ve^{\a\b}\psi_\b~, \qquad \psi_{\a}=\ve_{\a\b}\psi^\b~.
\eea

We use a Majorana representation in which all the $\g$-matrices are real
and any Majorana spinor $\j^\a$ is real,  
\bea
(\psi^{\a})^* = \psi^{\a} ~,\qquad (\psi_{\a})^*= \psi_{\a}  ~.
\eea
In such a representation the $3D$ gamma matrices $ (\g_a )_{\a  \b} $ and $ (\g_a )^{\a  \b} $ are {\it real} 
and symmetric. 

The matrices
\be
\g_a:=(\g_a)_\a{}^{\b}=\ve^{\b\g}(\g_a)_{\a\g}
\ee
satisfy the relations
\bsubeq
\bea
&\{\g_a,\g_b\}=2\eta_{ab}{\mathbbm 1}~,
\\
&\g_a\g_b=\eta_{ab}{\mathbbm 1}+\ve_{abc}\g^c~,
\eea
\esubeq
where the $3D$ Minkowski metric is $\eta_{ab}=\eta^{ab}={\rm diag}(-1,1,1)$ 
and the Levi-Civita tensor is normalized as $\ve_{012}=-\ve^{012}=-1$.
Some useful relations involving $\g$-matrices and the Levi-Civita tensor are 
\bsubeq
\bea
(\g^a)_{\a \b} (\g_a)_{\g \d} &=& 2\ve_{\a(\g}  \ve_{\d)\b} ~,
\\
\ve_{abc}(\g^b)_{\a\b}(\g^c)_{\g\d}&=&
\ve_{\g(\a}(\g_a)_{\b)\d}
+\ve_{\d(\a}(\g_a)_{\b)\g}
~,
\\
\tr[\g_a\g_b\g_{c}\g_d]&=&
2\eta_{ab}\eta_{cd}
-2\eta_{ac}\eta_{db}
+2\eta_{ad}\eta_{bc}
~,
\\
\eps_{abc} \eps^{def} &=& - 6 \d^d_{[a} \d^e_{b} \d^f_{c]}
~.
\eea
\esubeq

Given a three-vector, $V_a$, it can equivalently be realized as a symmetric spinor 
$V_{\a\b} =V_{\b \a}$.
The relationship between $V_a$ and $V_{\a \b}$ is as follows:
\bea
V_{\a\b}:=(\g^a)_{\a\b}V_a=V_{\b\a}~,\qquad
V_a=-\hf(\g_a)^{\a\b}V_{\a\b}~.
\label{vector-rule}
\eea
In three dimensions an
antisymmetric tensor $F_{ab}=-F_{ba}$ is Hodge-dual to a three-vector $F_a$:
\bea
F_a=\hf\ve_{abc}F^{bc}~,\qquad
F_{ab}=-\ve_{abc}F^c~.
\label{hodge-1}
\eea
The symmetric spinor $F_{\a\b} =F_{\b\a}$ associated with $F_a$, can 
equivalently be defined in terms of  $F_{ab}$: 
\bea
F_{\a\b}:=(\g^a)_{\a\b}F_a=\hf(\g^a)_{\a\b}\ve_{abc}F^{bc}
~.
\label{hodge-2}
\eea
It follows that the three algebraic objects, $F_a$, $F_{ab}$ and $F_{\a \b}$, 
are in one-to-one correspondence with each other, 
$F_a \leftrightarrow F_{ab} \leftrightarrow F_{\a\b}$.
Their corresponding inner products are related to each other as follows:
\bea
-F^aG_a=
\hf F^{ab}G_{ab}=\hf F^{\a\b}G_{\a\b}
~.
\eea

The spinor covariant derivatives in Minkowski superspace $\dsR^{3| 2 \cN}$ satisfy the anti-commutation relations
\be \{ D_\a^I , D_\b^J \} = 2 \ri \d^{IJ} (\g^c)_{\a\b} \partial_c \ .
\label{B.1}
\ee
They may be realized as
\be D_\a^I = \frac{\partial}{\partial \theta^\a_I} + \ri (\g^b)_{\a\b} \theta^\b_I \partial_b \ ,
\label{B.2}
\ee
where $\rm SO(\cN)$ indices may be equivalently written in the upper or lower position.

Due to the reality of the $3D$ spinors we have the conjugation rule
\be (D^I_\a F)^* = - (-1)^{\eps(F)} D^I_\a \bar{F} \ , \label{CDconjR}
\ee
with $F$ a superfield of
Grassmann parity $\eps(F)$ and $\bar{F} = (F)^*$.

For $\cN > 1$ it is useful to introduce the totally antisymmetric invariant $\rm SO(\cN)$ tensor
\be \eps^{I_1 \cdots I_{\cN}} = \eps^{[I_1 \cdots I_{\cN}]} \ ,
\ee
normalized as
\be \eps^{1 2 \cdots \cN} = \eps_{1 2 \cdots \cN} = 1 \ .
\ee
In the $\cN = 2$ case $\eps^{IJ}$ should not be confused with $\eps_{\a\b}$ since it is normalized differently.


\section{Coupling to a vector multiplet} \label{VM}

Here we do not consider general matter couplings within our formulation, however the constraints imposed on the geometry of section 
\ref{confSUGRA} were modeled on an abelian vector multiplet. It is therefore natural to discuss the coupling of an Abelian $\cN$-extended vector 
multiplet
\be V={\rm d}z^M V_M = E^A V_A \ , \quad V_A :=E_A{}^M V_M
\ee
to conformal supergravity, both for completeness and as a straightforward 
extension of the results in \cite{KLT-M11}. To do so we introduce the gauge covariant 
derivatives
\bea
{\bm\nabla}_A:=\nabla_A-V_A \bm Z
~, \qquad [\bm Z, \nabla_A ]=0~,
\label{2.23}
\eea
with $V_A(z)$ the gauge connection
associated with the generator $\bm Z$.
The gauge transformation of $V_A $ is 
\bea
\d
V_A = \nabla_A \t~,
\eea
with $\t (z)$ an arbitrary scalar superfield.

The algebra of covariant derivatives is found to be
\begin{align}
[{\bm \nabla}_A, {\bm \nabla}_B\} &= -T_{AB}{}^C{\bm \nabla}_C
-\hf \RM_{AB}{}^{cd} M_{cd}
-\hf \RN_{AB}{}^{PQ} N_{PQ}
- \RD_{AB} \mathbb D \non\\
&\quad - \RS_{AB}{}^\g_I S_\g^I
	- \RK_{AB}{}^c K_c
	-F_{AB}{\bm Z}
~,
\end{align}
where the torsion and curvatures are those of conformal superspace but with $F_{AB}$ the gauge-invariant field strength.
The field strength $F_{AB}$ satisfies the Bianchi identity
\bea
\nabla_{[A}F_{BC\}}+T_{[AB}{}^{D}F_{|D|C\}}=0~,
\eea
and must be subject to covariant constraints to describe an irreducible vector multiplet. The structure of the constraints 
and their consequence is different for $\cN=1$ and for $\cN >1$.

\subsection{The $\cN = 1$ case}

In the $\cN=1$ case, 
one imposes the covariant constraint \cite{Siegel,GGRS}
\bea
F_{\a\b}=0~.
\eea
Then one derives from the Bianchi identities the remaining components
\bsubeq
\bea
F_{a\b}&=&\hf (\g_a)_\b{}^\g G_\g
~,\\
F_{ab}&=&-\frac{\ri}{4}\ve_{abc}(\g^c)^{\g\d}\nabla_\g G_\d
~,
\eea
\esubeq
together with the dimension-2 differential constraint on the spinor field strength 
\bea
\nabla^\a G_\a=0~.
\eea
Furthermore, the Jacobi identities require $G_\a$ to be primary and of dimension-3/2:
\be S_\b G_\a = 0 \ , \quad \mathbb D G_\a = \frac{3}{2} G_\a \ .
\ee

\subsection{The $\cN > 1$ case}

For  $\cN>1$ one imposes the following dimension-1 covariant  constraint 
\cite{HitchinKLR,ZP,ZH}
\be
F_{\a}^I{}_\b^J =  -2\ri\ve_{\a\b}G^{IJ} \ ,
\ee
where $G^{IJ}$ is antisymmetric, primary and of dimension-$1$
\be
G^{IJ}=-G^{JI}~, \quad S_\a^I G^{JK} = 0 \ , \quad \mathbb D G^{IJ} = G^{IJ} \ .
\ee
Note that these constraints are a natural generalization of the 
$\cN>1$ constraints in four dimensions \cite{GSW,Sohnius}.
The Bianchi identities then give the remaining field strength components:
\bsubeq
\bea
F_{a}{}_\a^I&=&
\frac{1}{ (\cN-1)}(\g_a)_\a{}^{\b}\nabla_{\b J} G^{I J}
~,
\\
F_{ab}&=&
-\frac{\ri}{ 4\cN(\cN-1)}\ve_{abc}(\g^c)^{\a\b}[\nabla_{\a}^{ K},\nabla_{\b}^{ L}] G_{ K L}
~.
\eea
\esubeq

The $\cN=2$ case is special because $G^{IJ}$ becomes proportional to the 
antisymmetric tensor $\ve^{IJ}$
\bea
G^{IJ}= \ve^{IJ} G~.
\eea
The components of $F_{AB}$ then become
\bsubeq
\bea
F_\a^I{}_\b^J&=& -2\ri\ve_{\a\b}\ve^{IJ}G
~,
\\
F_a{}_\b^J&=&\ve^{JK}(\g_a)_\b{}^\g \nabla_{\g K}G
~,
\\
F_{ab}&=&
-\frac{\ri}{4} \ve_{abc}
(\g^c)^{\g\d}\ve^{KL}\nabla_{\g K} \nabla_{\d L} G~.
\eea
\esubeq
The Bianchi identities imply a constraint on $G$ at dimension-2
\bea
\ve^{K(I}\nabla^{\g J)} \nabla_{\g K} G=0 ~.
\label{2.32}
\eea
In the complex basis, this constraint means that $G$ is covariantly linear, 
\bea
\nabla^2 G = \bar \nabla^2 G=0~.
\eea

Unlike for $\cN = 2$, in the case  $\cN>2$ the field strength $G^{I J}$ is constrained by the 
dimension-3/2 Bianchi identity 
\bea
\nabla_{\g}^{I} G^{ J K}&=&
\nabla_{\g}^{[I} G^{ J K]}
- \frac{1}{ (\cN-1)}\big(\d^{I J}\nabla_{\g L} G^{ K L}
-\d^{I K}\nabla_{\g L} G^{ J L}\big)
~.
\label{2.35}
\eea
This constraint may be shown to define an off-shell supermultiplet 
\cite{GGHN}.\footnote{It was claimed in \cite{KLT-M11} 
that the constraint \eqref{2.35} defines an on-shell vector supermultiplet 
for $\cN>4$. This claim had been based on a harmonic-superspace
analysis by one of us (SMK), which turned out to be erroneous.}  
This is in contrast with the four-dimensional case where the standard superspace constraints 
define an on-shell vector multiplet for $\cN>2$ \cite{Sohnius}.

It is worth remarking briefly on why this difference should arise
between the three and four-dimensional cases. 
Following Sohnius \cite{Sohnius},
in four-dimensional $\cN$-extended Minkowski superspace,
the Abelian vector multiplet is described  by 
the complex 
field strength $\bar W^{jk} = - \bar W^{kj}$ with $\rm SU(\cN)$ indices. The field strength 
obeys the constraints 
\bsubeq
\begin{align}
D_\alpha^i \bar W^{jk} &= -D_\alpha^j \bar W^{ik}~, \\
\bar D_{\dalpha i} \bar W^{jk} &= \frac{1}{\cN-1} \Big(
	\delta_i^j \bar D_{\dalpha l} \bar W^{lk}
	- \delta_i^k \bar D_{\dalpha l} \bar W^{lj}
	\Big)~,
\end{align}
\esubeq
which (one can check) are conformally invariant. As a consequence
of these constraints, one can show for $\cN>2$ that
\begin{align}\label{eq:4Donshell}
\Box \bar W^{jk} = 0~,
\end{align}
which places the multiplet on-shell \cite{Sohnius}. Since the original constraints
are conformally invariant, any equation derived from them must also be
conformally invariant or transform under special conformal
transformations back into the original constraints. One easily
observes, for example, that eq. \eqref{eq:4Donshell} is invariant under $K_a$ 
because $\bar W^{jk}$ has conformal dimension 1.

In the three-dimensional case, one might expect that one could similarly
prove $\Box G^{JK} = 0$ for $\cN>4$. But in three dimensions, the 
massless Klein-Gordon
equation is conformally invariant only for Lorentz scalars of dimension-1/2.
Since $G^{JK}$ has dimension 1, one can prove that under successive applications of $K_a$,
\begin{align}
\Box G^{JK} = 0 \quad \overset{K_a\;\;\;}{\Longrightarrow} \quad
\nabla_a G^{JK} = 0 \quad \overset{K_a\;\;\;}{\Longrightarrow} \quad
G^{JK} = 0~.
\end{align}
In other words, provided superconformal invariance is maintained,
$\Box G^{JK}$ can vanish only if $G^{JK}$ also vanishes.

Similar arguments may be used to argue that a linearized version of the constraint
\eqref{CSBIN>3} defines an off-shell supermultiplet for $\cN>4$. This is in agreement
with the statement \cite{HIPT} that the $\cN$-extended Weyl multiplet is off-shell in 
three dimensions. 


\begin{footnotesize}

\end{footnotesize}


\begin{thebibliography}{66}
  
\bibitem{DK} 
S.~Deser and J.~H.~Kay,
``Topologically massive supergravity,''
Phys.\ Lett.\ B {\bf 120}, 97 (1983).

\bibitem{Deser}
  S.~Deser,
  ``Cosmological topological supergravity,''
 in {\it Quantum Theory Of Gravity}, S. M. Christensen (Ed.), 
 Adam Hilger, Bristol, 1984, pp. 374-381. 

\bibitem{vanN85}
P.~van Nieuwenhuizen,
``D = 3 conformal supergravity and Chern-Simons terms,''
Phys.\ Rev.\  D {\bf 32}, 872 (1985).
  
\bibitem{RvanN86} 
  M.~Ro\v{c}ek and P.~van Nieuwenhuizen,
  ``N $\geq$ 2 supersymmetric Chern-Simons terms as D = 3 extended conformal supergravity,''
  Class.\ Quant.\ Grav.\  {\bf 3}, 43 (1986).
  
\bibitem{LR89}
  U.~Lindstr\"om and M.~Ro\v{c}ek,
  ``Superconformal gravity in three dimensions as a gauge theory,''
  Phys.\ Rev.\ Lett.\  {\bf 62}, 2905 (1989).

\bibitem{NG}
  H.~Nishino and S.~J.~Gates, Jr.,
  ``Chern-Simons theories with supersymmetries in three dimensions,''
  Int.\ J.\ Mod.\ Phys.\  A {\bf 8}, 3371 (1993).

\bibitem{Uematsu}
  T.~Uematsu,
  ``Structure of N=1 conformal and Poincare supergravity in (1+1)-dimensions
  and (2+1)-dimensions,''
  Z.\ Phys.\  C {\bf 29}, 143 (1985);
``Constraints and actions in two-dimensional and three-dimensional N=1
conformal supergravity,''
Z.\ Phys.\  C {\bf 32}, 33 (1986).
 
\bibitem{BG}
  M.~Brown and S.~J.~Gates, Jr.,
``Superspace Bianchi identities and the supercovariant derivative,''
  Annals Phys.\  {\bf 122}, 443 (1979).

\bibitem{GGRS}
 S.~J.~Gates, Jr., M.~T.~Grisaru, M.~Ro\v{c}ek and W.~Siegel,
{\it Superspace, or One Thousand and One Lessons in Supersymmetry},
Front.\ Phys.\  {\bf 58}, 1 (1983) [arXiv:hep-th/0108200].

\bibitem{ZP}
  B.~M.~Zupnik and D.~G.~Pak,
  ``Superfield formulation of the simplest three-dimensional gauge theories and
  conformal supergravities,''  Theor.\ Math.\ Phys.\  {\bf 77} (1988) 1070
  [Teor.\ Mat.\ Fiz.\  {\bf 77} (1988) 97].
  
\bibitem{ZP89} 
B.~M.~Zupnik and D.~G.~Pak,
``Differential and integral forms in supergauge theories and supergravity,''
Class.\ Quant.\ Grav.\  {\bf 6}, 723 (1989).
  
 
 
\bibitem{HIPT}
P.~S.~Howe, J.~M.~Izquierdo, G.~Papadopoulos and P.~K.~Townsend,
``New supergravities with central charges and Killing spinors in 2+1 dimensions,''
Nucl.\ Phys.\  B {\bf 467}, 183 (1996)
  [arXiv:hep-th/9505032].

 \bibitem{Howe}
P.~S.~Howe,
``A superspace approach to extended conformal supergravity,''
Phys.\ Lett.\  B {\bf 100}, 389 (1981);
``Supergravity in superspace,''  Nucl.\ Phys.\  B {\bf 199}, 309 (1982).

  \bibitem{KLT-M11} 
  S.~M.~Kuzenko, U.~Lindstr\"om and G.~Tartaglino-Mazzucchelli,
  ``Off-shell supergravity-matter couplings in three dimensions,''
  JHEP {\bf 1103}, 120 (2011)
  [arXiv:1101.4013 [hep-th]].
  
 \bibitem{Howe:2004ib}
P.~S.~Howe and E.~Sezgin,
``The supermembrane revisited,''
Class.\ Quant.\ Grav.\ {\bf 22} (2005) 2167
[arXiv:hep-th/0412245].


\bibitem{CGN} 
  M.~Cederwall, U.~Gran and B.~E.~W.~Nilsson,
  ``D=3, N=8 conformal supergravity and the Dragon window,''
  JHEP {\bf 1109}, 101 (2011)
  [arXiv:1103.4530 [hep-th]].
  
\bibitem{GH}
 J.~Greitz and P.~S.~Howe,
``Maximal supergravity in three dimensions: supergeometry and differential forms,''
  JHEP {\bf 1107}, 071 (2011)
  [arXiv:1103.2730 [hep-th]].
  
\bibitem{KT-M11}
S.~M.~Kuzenko and G.~Tartaglino-Mazzucchelli,
  ``Three-dimensional N=2 (AdS) supergravity and associated supercurrents,''
JHEP {\bf 1112}, 052 (2011)
[arXiv:1109.0496 [hep-th]].

  
\bibitem{KT-M12} 
 S.~M.~Kuzenko and G.~Tartaglino-Mazzucchelli,
 ``Conformal supergravities as Chern-Simons theories revisited,''
JHEP {\bf 1303}, 113 (2013)
  [arXiv:1212.6852 [hep-th]].  

\bibitem{BKN} 
D.~Butter, S.~M.~Kuzenko and J.~Novak,
``The linear multiplet and ectoplasm,''
  JHEP {\bf 1209}, 131 (2012)
  [arXiv:1205.6981 [hep-th]].

\bibitem{Hasler} 
M.~F.~Hasler, ``The three-form multiplet in N=2 superspace,''
Eur.\ Phys.\ J.\ C {\bf 1}, 729 (1998) [hep-th/9606076].

\bibitem{Ectoplasm} 
S.~J.~Gates, Jr., ``Ectoplasm has no topology: The prelude,''
in {\it Supersymmetries and Quantum Symmetries},
 J. Wess and E. A. Ivanov (Eds.), Springer, Berlin, 1999, p. 46, arXiv:hep-th/9709104;
``Ectoplasm has no topology,''
 Nucl.\ Phys.\  B {\bf 541}, 615 (1999)
 [arXiv:hep-th/9809056].

\bibitem{GGKS}
S.~J.~Gates, Jr., M.~T.~Grisaru, M.~E.~Knutt-Wehlau and W.~Siegel,
``Component actions from curved superspace: Normal coordinates and
ectoplasm,'' Phys.\ Lett.\  B {\bf 421}, 203 (1998)
[hep-th/9711151].

\bibitem{ButterN=1} 
  D.~Butter, ``N=1 Conformal superspace in four dimensions,''
  Annals Phys.\  {\bf 325}, 1026 (2010)  [arXiv:0906.4399 [hep-th]].
  
\bibitem{ButterN=2} 
  D.~Butter,  ``N=2 Conformal superspace in four dimensions,''
JHEP {\bf 1110}, 030 (2011)  [arXiv:1103.5914 [hep-th]].  
  
\bibitem{FradTsey} 
  E.~S.~Fradkin and A.~A.~Tseytlin,
  ``Conformal supergravity,''
  Phys.\ Rept.\  {\bf 119}, 233 (1985).

\bibitem{FVP} 
  D.~Z.~Freedman and A.~Van Proeyen,
 {\it Supergravity},
  Cambridge, UK: Cambridge Univ. Pr. (2012) 607 p.
  
\bibitem{Nahm}  W.~Nahm,
``Supersymmetries and their representations,''
Nucl.\ Phys.\ B {\bf 135}, 149 (1978) .

\bibitem{KPT-MvU}
S.~M.~Kuzenko, J.~-H.~Park, G.~Tartaglino-Mazzucchelli and R.~von Unge,
``Off-shell superconformal nonlinear sigma-models in three dimensions,''
JHEP {\bf 1101}, 146 (2011)
 [arXiv:1011.5727 [hep-th]].

   
\bibitem{Kuzenko12} 
S.~M.~Kuzenko,
``Prepotentials for N=2 conformal supergravity in three dimensions,''
JHEP {\bf 1212}, 021 (2012)  [arXiv:1209.3894 [hep-th]].

\bibitem{BKNT-M}
D.~Butter, S.~M.~Kuzenko, J.~Novak and G.~Tartaglino-Mazzucchelli,
``Conformal supergravity in three dimensions: Off-shell actions,''
arXiv:1306.1205 [hep-th]. 



\bibitem{GIOS}
A.~S.~Galperin, E.~A.~Ivanov, V.~I.~Ogievetsky and E.~S.~Sokatchev,
{\it Harmonic Superspace}, Cambridge University Press, 
Cambridge, 2001.



\bibitem{KT}
S.~M.~Kuzenko and S.~Theisen,
 ``Correlation functions of conserved currents in N = 2 superconformal
theory,''  Class.\ Quant.\ Grav.\  {\bf 17}, 665 (2000)  [hep-th/9907107]. 


\bibitem{Siegel-curved}
  W.~Siegel,
  ``Curved extended superspace from Yang-Mills theory a la strings,''
  Phys.\ Rev.\  D {\bf 53}, 3324 (1996)
  [hep-th/9510150].

\bibitem{NT} 
  M.~Nishimura and Y.~Tanii,
  ``Coupling of the BLG theory to a conformal supergravity background,''
  JHEP {\bf 1301}, 120 (2013)
  [arXiv:1206.5388 [hep-th]].

\bibitem{deWN} 
  B.~de Wit and H.~Nicolai,
 ``N=8 supergravity,''
  Nucl.\ Phys.\ B {\bf 208}, 323 (1982).
  
\bibitem{WB}
J.~Wess and J.~Bagger,
{\it Supersymmetry and Supergravity},
Princeton University Press, Princeton, 1992.

\bibitem{Ideas} 
I.~L. Buchbinder and S.~M. Kuzenko, {\it Ideas and Methods of Supersymmetry and
Supergravity, Or a Walk Through Superspace}, IOP, Bristol, 1998.

\bibitem{Siegel}
  W.~Siegel,  ``Unextended superfields in extended supersymmetry,''
  Nucl.\ Phys.\  B {\bf 156}, 135 (1979).
  
\bibitem{HitchinKLR}
N.~J.~Hitchin, A.~Karlhede, U.~Lindstr\"om and M.~Ro\v cek,
``Hyperk\"ahler metrics and supersymmetry,''
Commun.\ Math.\ Phys.\  {\bf 108}, 535 (1987).  
  
\bibitem{ZH}
B.~M.~Zupnik and D.~V.~Hetselius,
``Three-dimensional extended supersymmetry in harmonic superspace,''
Sov.\ J.\ Nucl.\ Phys.\  {\bf 47}, 730 (1988)
  [Yad.\ Fiz.\  {\bf 47}, 1147 (1988)].

\bibitem{GSW}
  R.~Grimm, M.~Sohnius and J.~Wess,
  ``Extended supersymmetry and gauge theories,''
  Nucl.\ Phys.\  B {\bf 133}, 275 (1978).
  
\bibitem{Sohnius}
  M.~F.~Sohnius,
``Bianchi identities for supersymmetric gauge theories,''
  Nucl.\ Phys.\  B {\bf 136}, 461 (1978).
  
  

\bibitem{GGHN} 
  U.~Gran, J.~Greitz, P.~Howe and B.~E.~W.~Nilsson,
  ``Topologically gauged superconformal Chern-Simons matter theories,''
  JHEP {\bf 1212}, 046 (2012)
  [arXiv:1204.2521 [hep-th]].
 


\end{thebibliography}
\end{document}